\pdfoutput=1

\documentclass[12pt,a4paper]{article}

\usepackage{ifthen} 
\newboolean{pdflatex}
\setboolean{pdflatex}{true} 

\newboolean{articletitles}
\setboolean{articletitles}{true}

\newboolean{uprightparticles}
\setboolean{uprightparticles}{false}

\def\paperauthors{LHCb collaboration} 
\def\paperasciititle{Branching fraction measurements of the rare Bs to phimumu and Bs to f_2'(1525)mumu decays} 
\def\papertitle{Branching fraction measurements\\ of the rare $\decay{\Bs}{\phi\mumu}$ and \\ $\decay{\Bs}{f_2^\prime(1525)\mumu}$ decays}
\def\paperkeywords{{High Energy Physics}, {LHCb}} 
\def\papercopyright{\the\year\ CERN for the benefit of the LHCb collaboration} 
\def\paperlicence{CC BY 4.0 licence}
\def\paperlicenceurl{https://creativecommons.org/licenses/by/4.0/}

\usepackage[top=1in, bottom=1.25in, left=1in, right=1in]{geometry}

\usepackage{microtype}
\usepackage{lineno}  
\usepackage{xspace} 
\usepackage{caption} 

\usepackage{graphicx}  
\usepackage{color}
\usepackage{colortbl}
\graphicspath{{./figs/}} 

\usepackage{amsmath} 
\usepackage{amssymb}
\usepackage{amsfonts}
\usepackage{upgreek} 

\newcommand*\patchAmsMathEnvironmentForLineno[1]{%
\expandafter\let\csname old#1\expandafter\endcsname\csname #1\endcsname
\expandafter\let\csname oldend#1\expandafter\endcsname\csname
end#1\endcsname
 \renewenvironment{#1}%
   {\linenomath\csname old#1\endcsname}%
   {\csname oldend#1\endcsname\endlinenomath}%
}
\newcommand*\patchBothAmsMathEnvironmentsForLineno[1]{%
  \patchAmsMathEnvironmentForLineno{#1}%
  \patchAmsMathEnvironmentForLineno{#1*}%
}
\AtBeginDocument{%
\patchBothAmsMathEnvironmentsForLineno{equation}%
\patchBothAmsMathEnvironmentsForLineno{align}%
\patchBothAmsMathEnvironmentsForLineno{flalign}%
\patchBothAmsMathEnvironmentsForLineno{alignat}%
\patchBothAmsMathEnvironmentsForLineno{gather}%
\patchBothAmsMathEnvironmentsForLineno{multline}%
\patchBothAmsMathEnvironmentsForLineno{eqnarray}%
}

\usepackage{hyperxmp}

\usepackage[pdftex,
            pdfauthor={\paperauthors},
            pdftitle={\paperasciititle},
            pdfkeywords={\paperkeywords},
            pdfcopyright={Copyright (C) \papercopyright},
            pdflicenseurl={\paperlicenceurl}]{hyperref}

\usepackage[colorinlistoftodos,textsize=scriptsize]{todonotes}

\usepackage[bottom,flushmargin,hang,multiple]{footmisc}

\usepackage[all]{hypcap} 

\usepackage{xspace} 
\usepackage{upgreek}

\def\lhcb   {\mbox{LHCb}\xspace}

\def\MagUp {\mbox{\em Mag\kern -0.05em Up}\xspace}

\ifthenelse{\boolean{uprightparticles}}%
{

 \def\Pmu         {\ensuremath{\upmu}\xspace}

 \def\Ppi         {\ensuremath{\uppi}\xspace}

 \def\Pphi        {\ensuremath{\upphi}\xspace}

 \def\Ppsi        {\ensuremath{\uppsi}\xspace}

 \def\PDelta      {\ensuremath{\Delta}\xspace}                 
 \def\PXi         {\ensuremath{\Xi}\xspace}                 
 \def\PLambda     {\ensuremath{\Lambda}\xspace}                 
 \def\PSigma      {\ensuremath{\Sigma}\xspace}                 
 \def\POmega      {\ensuremath{\Omega}\xspace}                 
 \def\PUpsilon    {\ensuremath{\Upsilon}\xspace}

 \def\PB      {\ensuremath{\mathrm{B}}\xspace}                 
                  
 \def\PD      {\ensuremath{\mathrm{D}}\xspace}

 \def\PJ      {\ensuremath{\mathrm{J}}\xspace}                 
 \def\PK      {\ensuremath{\mathrm{K}}\xspace}

 \def\Pb      {\ensuremath{\mathrm{b}}\xspace}

 \def\Pi      {\ensuremath{\mathrm{i}}\xspace}

 \def\Ps      {\ensuremath{\mathrm{s}}\xspace}

 \def\thebaroffset{0.0em}
}
{

 \def\Pmu         {\ensuremath{\mu}\xspace}

 \def\Ppi         {\ensuremath{\pi}\xspace}

 \def\Pphi        {\ensuremath{\phi}\xspace}

 \def\Ppsi        {\ensuremath{\psi}\xspace}                 
                  
 \mathchardef\PDelta="7101
 \mathchardef\PXi="7104
 \mathchardef\PLambda="7103
 \mathchardef\PSigma="7106
 \mathchardef\POmega="710A
 \mathchardef\PUpsilon="7107
                  
 \def\PB      {\ensuremath{B}\xspace}                 
                  
 \def\PD      {\ensuremath{D}\xspace}

 \def\PJ      {\ensuremath{J}\xspace}                 
 \def\PK      {\ensuremath{K}\xspace}

 \def\Pb      {\ensuremath{b}\xspace}

 \def\Pi      {\ensuremath{i}\xspace}

 \def\Ps      {\ensuremath{s}\xspace}

 \def\thebaroffset{0.18em}
}
\newcommand{\offsetoverline}[2][\thebaroffset]{\kern #1\overline{\kern -#1 #2}}%

\makeatletter
\ifcase \@ptsize \relax
  \newcommand{\miniscule}{\@setfontsize\miniscule{4}{5}}
\or
  \newcommand{\miniscule}{\@setfontsize\miniscule{5}{6}}
\or
  \newcommand{\miniscule}{\@setfontsize\miniscule{5}{6}}
\fi
\makeatother

\DeclareRobustCommand{\optbar}[1]{\shortstack{{\miniscule (\rule[.5ex]{1.25em}{.18mm})}
  \\ [-.7ex] $#1$}}

\def\mup        {{\ensuremath{\Pmu^+}}\xspace}
\def\mun        {{\ensuremath{\Pmu^-}}\xspace} 

\def\mumu       {{\ensuremath{\Pmu^+\Pmu^-}}\xspace}

\def\ellell     {\ensuremath{\ell^+ \ell^-}\xspace}

\def\squark    {{\ensuremath{\Ps}}\xspace}

\def\bquark    {{\ensuremath{\Pb}}\xspace}

\def\pion   {{\ensuremath{\Ppi}}\xspace}

\def\pim    {{\ensuremath{\pion^-}}\xspace}

\def\kaon    {{\ensuremath{\PK}}\xspace}

\def\KorKbar {\kern \thebaroffset\optbar{\kern -\thebaroffset \PK}{}\xspace}

\def\Kp      {{\ensuremath{\kaon^+}}\xspace}
\def\Km      {{\ensuremath{\kaon^-}}\xspace}

\def\Kstarz  {{\ensuremath{\kaon^{*0}}}\xspace}

\def\Kstar   {{\ensuremath{\kaon^*}}\xspace}

\newcommand{\phiz}{\ensuremath{\Pphi}\xspace}

\def\D       {{\ensuremath{\PD}}\xspace}

\def\DorDbar {\kern \thebaroffset\optbar{\kern -\thebaroffset \PD}\xspace}

\def\Dp      {{\ensuremath{\D^+}}\xspace}
\def\Dm      {{\ensuremath{\D^-}}\xspace}

\def\DpDm    {\ensuremath{\Dp {\kern -0.16em \Dm}}\xspace}

\def\B       {{\ensuremath{\PB}}\xspace}
\def\Bbar    {{\ensuremath{\offsetoverline{\PB}}}\xspace}

\def\BorBbar {\kern \thebaroffset\optbar{\kern -\thebaroffset \PB}\xspace}
\def\Bz      {{\ensuremath{\B^0}}\xspace}

\def\Bd      {{\ensuremath{\B^0}}\xspace}

\def\BdorBdbar {\kern \thebaroffset\optbar{\kern -\thebaroffset \Bd}\xspace}
\def\Bu      {{\ensuremath{\B^+}}\xspace}

\def\Bp      {{\ensuremath{\Bu}}\xspace}

\def\Bs      {{\ensuremath{\B^0_\squark}}\xspace}
\def\Bsb     {{\ensuremath{\Bbar{}^0_\squark}}\xspace}
\def\BsorBsbar {\kern \thebaroffset\optbar{\kern -\thebaroffset \Bs}\xspace}

\def\jpsi     {{\ensuremath{{\PJ\mskip -3mu/\mskip -2mu\Ppsi}}}\xspace}
\def\psitwos  {{\ensuremath{\Ppsi{(2S)}}}\xspace}

\def\Y#1S{\ensuremath{\PUpsilon{(#1S)}}\xspace}

\def\Lz          {{\ensuremath{\PLambda}}\xspace}

\def\LorLbar     {\kern \thebaroffset\optbar{\kern -\thebaroffset \PLambda}\xspace}

\def\Lb           {{\ensuremath{\Lz^0_\bquark}}\xspace}

\newcommand{\decay}[2]{\ensuremath{#1\!\to #2}\xspace} 

\def\to                 {\ensuremath{\rightarrow}\xspace}

\def\qsq       {{\ensuremath{q^2}}\xspace}

\def\CP                {{\ensuremath{C\!P}}\xspace}

\def\AT#1     {\ensuremath{A_{\mathrm{T}}^{#1}}\xspace}           

\def\C#1      {\ensuremath{\mathcal{C}_{#1}}\xspace}                       
\def\Cp#1     {\ensuremath{\mathcal{C}_{#1}^{'}}\xspace}                    
\def\Ceff#1   {\ensuremath{\mathcal{C}_{#1}^{\mathrm{(eff)}}}\xspace}        
\def\Cpeff#1  {\ensuremath{\mathcal{C}_{#1}^{'\mathrm{(eff)}}}\xspace}       
\def\Ope#1    {\ensuremath{\mathcal{O}_{#1}}\xspace}                       
\def\Opep#1   {\ensuremath{\mathcal{O}_{#1}^{'}}\xspace}                    

\newcommand{\aunit}[1]{\ensuremath{\text{\,#1}}}       

\newcommand{\tev}{\aunit{Te\kern -0.1em V}\xspace}
\newcommand{\gev}{\aunit{Ge\kern -0.1em V}\xspace}
\newcommand{\mev}{\aunit{Me\kern -0.1em V}\xspace}
\newcommand{\kev}{\aunit{ke\kern -0.1em V}\xspace}
\newcommand{\ev}{\aunit{e\kern -0.1em V}\xspace}
 
\newcommand{\mevc}{\ensuremath{\aunit{Me\kern -0.1em V\!/}c}\xspace}
\newcommand{\gevc}{\ensuremath{\aunit{Ge\kern -0.1em V\!/}c}\xspace}
\newcommand{\mevcc}{\ensuremath{\aunit{Me\kern -0.1em V\!/}c^2}\xspace}
\newcommand{\gevcc}{\ensuremath{\aunit{Ge\kern -0.1em V\!/}c^2}\xspace}
\newcommand{\gevgevcccc}{\ensuremath{\gev^2\!/c^4}\xspace} 

\def\fb   {\ensuremath{\aunit{fb}}\xspace}
\def\invfb   {\ensuremath{\fb^{-1}}\xspace}

\def\gsim{{~\raise.15em\hbox{$>$}\kern-.85em
          \lower.35em\hbox{$\sim$}~}\xspace}
\def\lsim{{~\raise.15em\hbox{$<$}\kern-.85em
          \lower.35em\hbox{$\sim$}~}\xspace}

\def\tell1  {TELL1\xspace}
\def\ukl1   {UKL1\xspace}

\newcommand{\eg}{\mbox{\itshape e.g.}\xspace}

\usepackage{cite} 
\usepackage{mciteplus}

\begin{document}

\renewcommand{\thefootnote}{\fnsymbol{footnote}}
\setcounter{footnote}{1}
\newcommand{\ftwo}{\ensuremath{f_2^\prime}}
\newcommand{\ftwoprime}{\ftwo}


\begin{titlepage}
\pagenumbering{roman}

\vspace*{-1.5cm}
\centerline{\large EUROPEAN ORGANIZATION FOR NUCLEAR RESEARCH (CERN)}
\vspace*{1.5cm}
\noindent
\begin{tabular*}{\linewidth}{lc@{\extracolsep{\fill}}r@{\extracolsep{0pt}}}
\ifthenelse{\boolean{pdflatex}}
{\vspace*{-1.5cm}\mbox{\!\!\!\includegraphics[width=.14\textwidth]{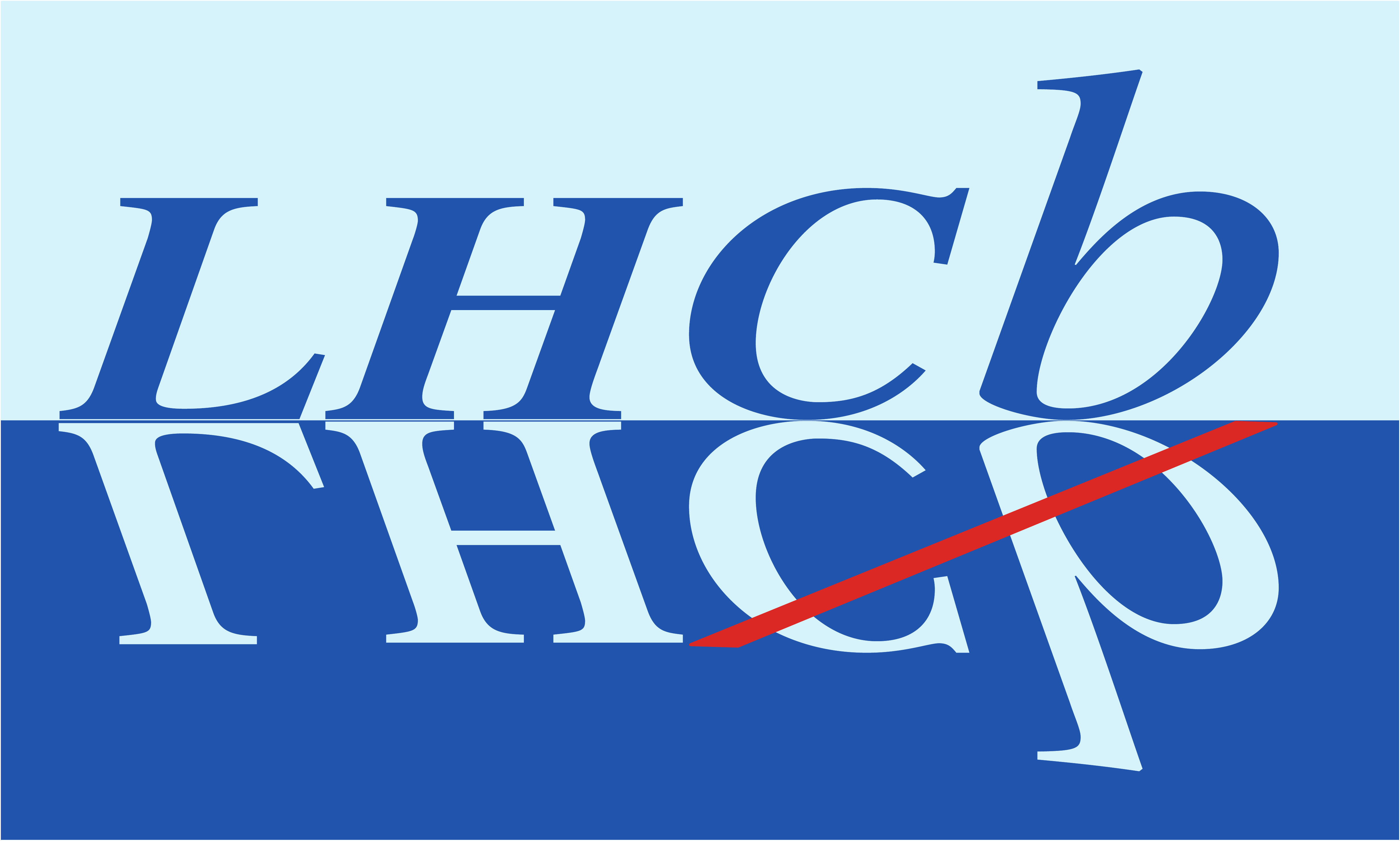}} & &}%
{\vspace*{-1.2cm}\mbox{\!\!\!\includegraphics[width=.12\textwidth]{figs/lhcb-logo.eps}} & &}%
\\
 & & CERN-EP-2021-092 \\  
 & & LHCb-PAPER-2021-014 \\  
 & & October 6, 2021 \\ 
 & & \\
\end{tabular*}

\vspace*{4.0cm}

{\normalfont\bfseries\boldmath\huge
\begin{center}
  \papertitle 
\end{center}
}

\vspace*{2.0cm}

\begin{center}
\paperauthors\footnote{Authors are listed at the end of this paper.}
\end{center}

\vspace{\fill}

\begin{abstract}
  \noindent
The branching fraction of the rare \mbox{$B^0_s\rightarrow\phi\mu^+\mu^-$} decay is measured using data collected by the \mbox{LHCb}\ experiment
at center-of-mass energies of 7, 8 and 13\,$\rm{TeV}$,
 corresponding to integrated luminosities of $1$, $2$ and $6\,{\rm fb}^{-1}$, respectively. 
The branching fraction is reported in intervals of $q^2$, the square of the dimuon invariant mass. 
In the $q^2$ region between $1.1$ and $6.0\aunit{Ge\kern -0.1em V}^2\!/c^4$, the measurement is found to lie $3.6$ standard deviations below a Standard Model prediction based on a combination of Light Cone Sum Rule and Lattice QCD calculations.
In addition, the first observation of the rare 
\mbox{$B^0_s\rightarrow f_2^\prime(1525)\mu^+\mu^-$} decay is reported with a statistical significance of nine standard deviations and its branching fraction is determined. 
\end{abstract}

\vspace*{2.0cm}

\begin{center}
  Published in
  Phys.~Rev.~Lett.~\textbf{127}~(2021)~151801 
\end{center}

\vspace{\fill}

{\footnotesize 
\centerline{\copyright~\papercopyright. \href{\paperlicenceurl}{\paperlicence}.}}
\vspace*{2mm}

\end{titlepage}


\newpage
\setcounter{page}{2}
\mbox{~}

\renewcommand{\thefootnote}{\arabic{footnote}}
\setcounter{footnote}{0}

\cleardoublepage

\pagestyle{plain} 
\setcounter{page}{1}
\pagenumbering{arabic}


Recent studies of rare semileptonic $\decay{b}{s\ell^+\ell^-}$ decays exhibit tensions between experimental results and Standard Model (SM) predictions of branching fractions~\cite{LHCb-PAPER-2015-023,LHCb-PAPER-2013-017,LHCb-PAPER-2016-012,LHCb-PAPER-2015-009,LHCb-PAPER-2014-006}, angular distributions~\cite{LHCb-PAPER-2020-002,LHCb-PAPER-2015-051,Aaboud:2018krd,Khachatryan:2015isa,Sirunyan:2017dhj,Wehle:2016yoi}, and lepton universality~\cite{LHCb-PAPER-2021-004,LHCb-PAPER-2019-040,LHCb-PAPER-2019-009,LHCb-PAPER-2017-013,LHCb-PAPER-2014-024,Wehle:2016yoi,Lees:2012tva,Abdesselam:2019lab,Abdesselam:2019wac}.
Since these decays are only allowed via higher-order electroweak (loop) diagrams in the SM, they constitute powerful probes for non-SM contributions. 
One of the most significant discrepancies 
appears in the branching fraction of the $\decay{\Bs}{\phi\mumu}$ decay~\cite{LHCb-PAPER-2015-023,LHCb-PAPER-2013-017}. 
Using 3\invfb of data collected with the LHCb experiment at center-of-mass energies of 7 and 8\tev, the branching fraction was measured below the SM prediction at the level of three standard deviations~($\sigma$)~\cite{LHCb-PAPER-2015-023}.  
This Letter presents an updated measurement 
using data taken at center-of-mass energies 
of 7, 8 and $13\tev$ during the 2011, 2012 and 2015--2018 data-taking periods, with integrated luminosities corresponding to $1$, $2$ and $6\invfb$, respectively. 
Compared to the 3\invfb sample alone, this represents an increase of about a factor of four in the number of produced \Bs\ mesons.
The branching fraction is determined in intervals of \qsq, the squared invariant mass of the dimuon system. 
In addition, the observation of the $\decay{\Bs}{f_2^\prime(1525)\mumu}$ decay and a determination of its branching fraction are reported. 
This constitutes the first observation of a rare semileptonic decay involving a spin-$2$ meson in the final state, and provides complementary information to
transitions involving pseudoscalar or vector mesons. 
In the following, the shorthand notation \ftwoprime\ is used to refer to the $f_2^\prime(1525)$ meson. The inclusion of charge-conjugate processes is implied throughout.

The LHCb detector is a single-arm forward spectrometer covering the pseudorapidity range $2<\eta<5$, detailed in Refs.~\cite{LHCb-DP-2008-001,LHCb-DP-2014-002}. 
The online event selection is performed by a trigger~\cite{LHCb-DP-2012-004} that consists of hardware and software stages. The former selects signal candidates containing a muon with significant transverse momentum with respect to the beam axis. 
At the software stage, a full event reconstruction is applied. 
Simulated events are used in this analysis to determine the reconstruction and selection efficiency of signal candidates,  
and to estimate contamination from residual background. 
The simulated samples are produced using the software described in
Refs.~\cite{Sjostrand:2007gs,Lange:2001uf,Allison:2006ve, *Agostinelli:2002hh}. 
Residual mismodeling in simulation is corrected for using control samples from data.  

The $\decay{\Bs}{\phi\mumu}$ and $\decay{\Bs}{\ftwoprime\mumu}$ decays are reconstructed in the $\Kp\Km\mumu$ final state. 
Particle identification criteria are applied to the kaon and muon candidates.
The muons (kaons) are further required to have $\chi^2_{\rm IP} > 9 (6)$ with respect to any primary $pp$ interaction vertex (PV) in the event, 
where $\chi^2_{\rm IP}$ denotes the difference in the vertex-fit $\chi^2$ of the PV 
when reconstructed with or without the considered track. 
The four final-state tracks are fit to a common vertex that is required to have good quality and to be significantly displaced from any PV in the event. 
Signal candidates are retained if the \Kp\Km\mumu invariant mass, $m(\Kp\Km\mumu)$, lies
between $5270$ and $5700\mevcc$. 
The invariant mass of the dikaon system, $m(\Kp\Km)$, is required to be within $12\mevcc$ of the known $\phi$ mass for the $\decay{\Bs}{\phi\mumu}$ decay, 
or within $225\mevcc$ of the known mass of the wider $\ftwoprime$ resonance for the $\decay{\Bs}{\ftwoprime\mumu}$ decay~\cite{pdg2020}.

The \qsq regions between $8.0$ and $11.0\gevgevcccc$, and between $12.5$ and $15.0\gevgevcccc$, are dominated by tree-level \Bs decays into final states with a $\jpsi$ or $\psitwos$ meson.
While these regions are vetoed in the selection of the signal modes, the decays to charmonium are used as high-yield control modes. The $\decay{\Bs}{\jpsi\phi}$ decay is used for normalization. 
The \qsq\ region from $0.98$ to $1.1\gevgevcccc$ is also vetoed to remove  contributions from $\decay{\Bs}{\phi(\to\mumu)\phi}$ decays.

To reduce combinatorial background, formed from random track combinations, a boosted decision tree (BDT) algorithm~\cite{Breiman,AdaBoost} is applied.
The BDT classifier is trained on data using cross-validation techniques~\cite{Blum:1999:BHB:307400.307439}, with $\decay{\Bs}{\jpsi\phi}$ events as signal proxy and candidates from the upper mass sideband $m(\Kp\Km\mumu)>5567\mevcc$ as background proxy. 
The classifier combines the \Bs\ transverse momentum and $\chi^2_{\rm IP}$, the angle between the \Bs\ momentum and the vector connecting the PV and the decay vertex of the \Bs candidate, 
the fit quality of the \Bs\ vertex and its displacement from the associated PV,  
particle identification information
and $\chi^2_{\rm IP}$ of 
the final-state particles.

The criterion on the BDT output is optimized by maximizing the expected significance of the $\decay{\Bs}{\phi\mumu}$ and $\decay{\Bs}{\ftwoprime\mumu}$ signals separately, due to different levels of background contamination. 
The requirement on the BDT classifier yields a signal 
efficiency of $96\%$ ($85\%$) and a background rejection of $96\%$ ($95\%$) for the \decay{\Bs}{\phi\mumu} \mbox{(\decay{\Bs}{\ftwoprime\mumu})} decay mode.
Finally, information from particle identification is combined with invariant mass variables, constructed under the relevant particle-hypotheses, to reject background from $\decay{\Lb}{p\Km\mumu}$ decays, where the proton is misidentified as a kaon,
and from $\decay{\Bs}{\jpsi\phi}$, $\decay{\Bs}{\psitwos\phi}$ and $\decay{\Bd}{\jpsi\Kstarz}$ decays, where a final state hadron is misreconstructed as a muon and vice-versa. 

The differential branching fraction of the $\decay{\Bs}{\phi\mumu}$ decay is determined in intervals of \qsq, relative to the $\decay{\Bs}{\jpsi\phi}$ normalization mode, according to
\begin{align}
    \frac{{\rm d}{\mathcal B}(\decay{\Bs}{\phi\mumu})}{{\rm d}\qsq} &= \frac{{\mathcal B}(\decay{\Bs}{\jpsi\phi})\times{\mathcal B}(\decay{\jpsi}{\mumu})}{q^2_{\rm max}-q^2_{\rm min}}\times \frac{N_{\phi\mumu}}{N_{\jpsi\phi}}\times \frac{\epsilon_{\jpsi\phi}}{\epsilon_{\phi\mumu}},\label{eq:diffbf}
\end{align}
where $N_{\jpsi\phi}$ and $\epsilon_{\jpsi\phi}$ are the yields and efficiencies of the normalization mode, and $N_{\phi\mumu}$ and $\epsilon_{\phi\mumu}$ the corresponding parameters for the signal mode in the $[q^2_{\rm min}, q^2_{\rm max}]$ interval.
The branching fractions related to the normalization mode are given by \mbox{${\mathcal B}(\decay{\Bs}{\jpsi\phi})=(1.018\pm 0.032\pm 0.037)\times 10^{-3}$~\cite{LHCb-PAPER-2020-046}} and \mbox{${\mathcal B}(\decay{\jpsi}{\mumu})=(5.961\pm 0.033)\,\%$~\cite{pdg2020}}.  

\begin{figure}
    \centering
    \includegraphics[width=7.5cm]{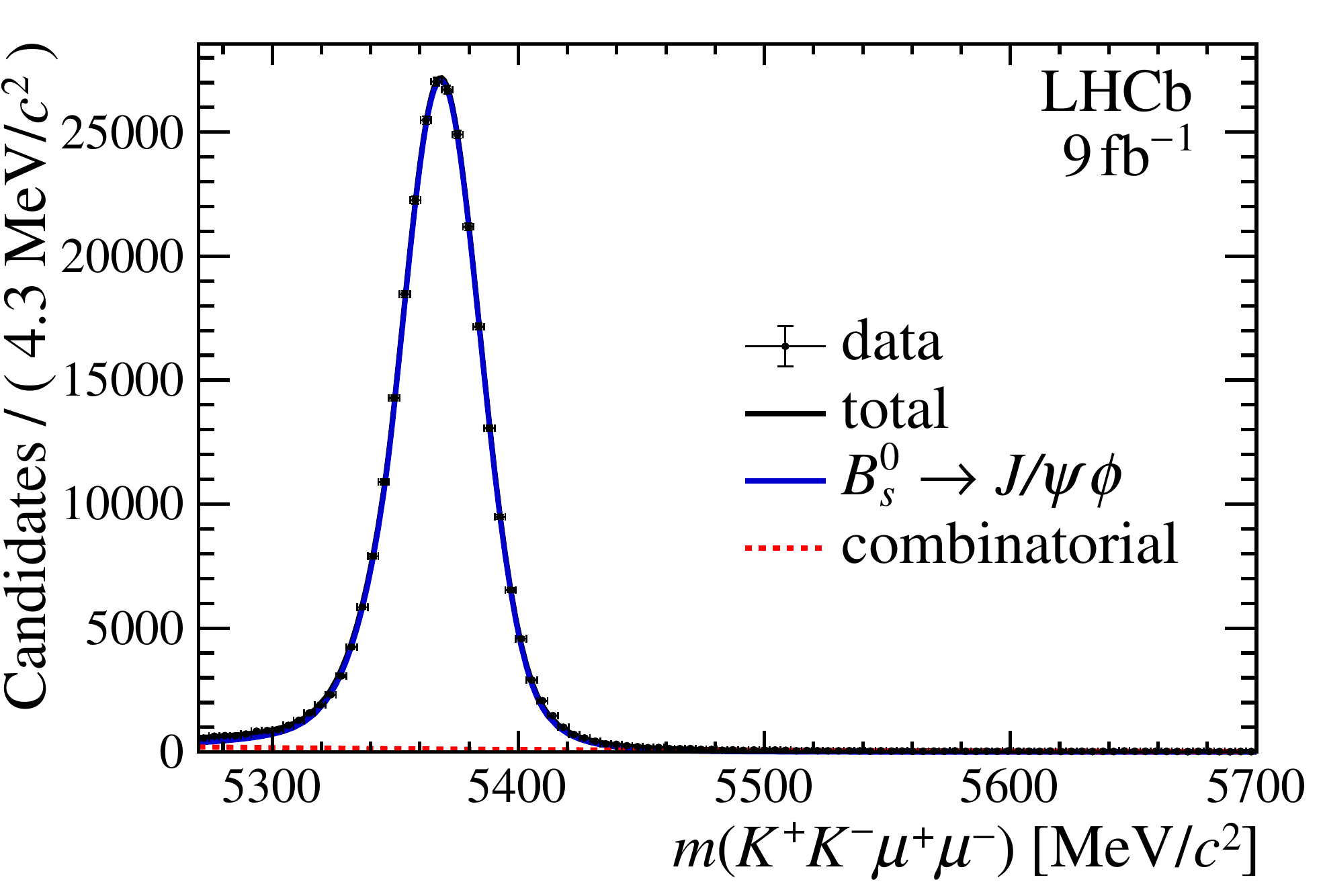}\hspace*{0.5cm}
    \includegraphics[width=7.5cm]{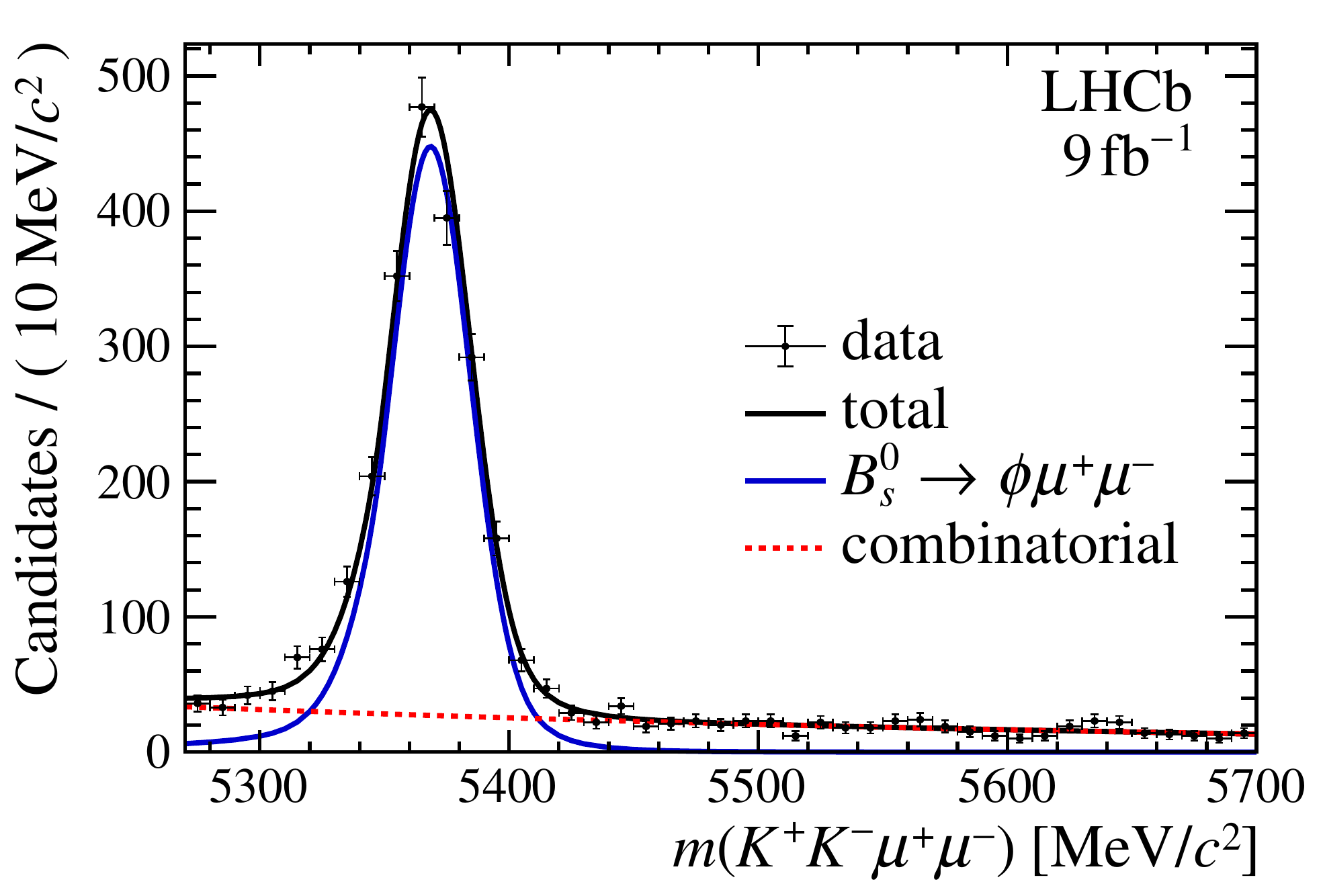}
    \caption{
        Reconstructed invariant mass of the $\Kp\Km\mup\mun$ system for (left) the $\decay{\Bs}{\jpsi\phi}$ normalization mode and (right) the $\decay{\Bs}{\phi\mumu}$ signal candidates, integrated over \qsq and overlaid with the fit projections. 
    }
    \label{fig:phiprojections}
\end{figure}

As the relative efficiencies vary according to 
the data-taking conditions, 
the data are split into the 2011--2012, 2015--2016 and 2017--2018 periods. 
The yields of the normalization mode for the different data-taking periods are determined using extended unbinned maximum-likelihood fits to the $m(\Kp\Km\mumu)$ distribution. 
The $\decay{\Bs}{\jpsi\phi}$ decay is modeled using 
the sum of two Gaussian functions with a common mean and a power-law tail towards upper and lower mass. The combinatorial background is modeled using an exponential function. 
The $m(\Kp\Km\mumu)$ distribution of the normalization mode for the full data sample, overlaid with 
the fit projections,
is shown in Fig.~\ref{fig:phiprojections}~(left). 
The yields of the normalization mode, $N_{\jpsi\phi}$, are determined to be $62\,980\pm270$, $70\,970\pm290$, and $148\,490\pm410$
for the three different data-taking periods, where the uncertainties are statistical only.

For the rare $\decay{\Bs}{\phi\mumu}$ decay, 
a simultaneous extended maximum-likelihood fit 
of the data samples for the different periods is performed in intervals of \qsq, 
where the signal yields are 
parameterized using Eq.~(\ref{eq:diffbf}) and the differential branching fraction is shared between the samples. 
The model used to describe the $m(\Kp\Km\mumu)$ distribution 
is the same as for
the $\decay{\Bs}{\jpsi\phi}$ normalization mode. 
The model parameters for the signal component are fixed to those from the fit of the normalization mode, where the \qsq\ dependence of the mass resolution is accounted for with scaling factors determined from simulation.

Negligible contributions from physical background, including $\decay{\Bs}{K^{+}K^{-}\mumu}$ decays with the $K^{+}K^{-}$ system in an S-wave configuration, are not considered in the fit and a systematic uncertainty is assigned. 
Integrated over the full \qsq\ range, signal yields, $N_{\phi\mumu}$, of \mbox{$458\pm 12$}, \mbox{$484\pm 13$}, and \mbox{$1064\pm 28$} are found
from the simultaneous fit to the different data sets. 
Figure~\ref{fig:phiprojections}~(right) shows the $m(\Kp\Km\mumu)$ distribution  
of the full data sample, integrated over \qsq and overlaid with the 
fit projections.
Figures for the different data-taking periods are available as Supplemental Material. 
 
The relative branching fraction measurement is affected by systematic uncertainties on the fit model and the efficiency ratio, where the latter is determined using SM simulation. 
A summary of the systematic uncertainties is provided in the Supplemental Material. 
The dominant systematic uncertainty on the absolute branching fraction (Eq.~\ref{eq:diffbf}) originates from the model used to simulate \decay{\Bs}{\phi\mumu} events ($0.04$--$0.10\times10^{-8}\gev^{-2}c^{4}$). 
The model depends on $\Delta\Gamma_s$, the decay width difference in the \Bs\ system~\cite{Descotes-Genon:2015hea}, and the 
specific form factors used.
The effect of the model-choice on the relative efficiency is assessed by varying $\Delta\Gamma_s$ by 20\%, corresponding to the difference in $\Delta\Gamma_s$ between the default value~\cite{PDG2014} and that of Ref.~\cite{pdg2020}, and by comparing the form factors in Ref.~\cite{Straub:2015ica} with the older calculations in Ref.~\cite{Ball:2004rg}. The observed differences are taken as a systematic uncertainty. Other leading sources of systematic uncertainty arise from the limited size of the simulation sample ($0.02$--$0.07\times10^{-8}\gev^{-2}c^{4}$) and the omission of small background contributions from the fit model ($0.01$--$0.04\times10^{-8}\gev^{-2}c^{4}$).

\begin{figure}
    \centering
    \includegraphics[width=9.5cm]{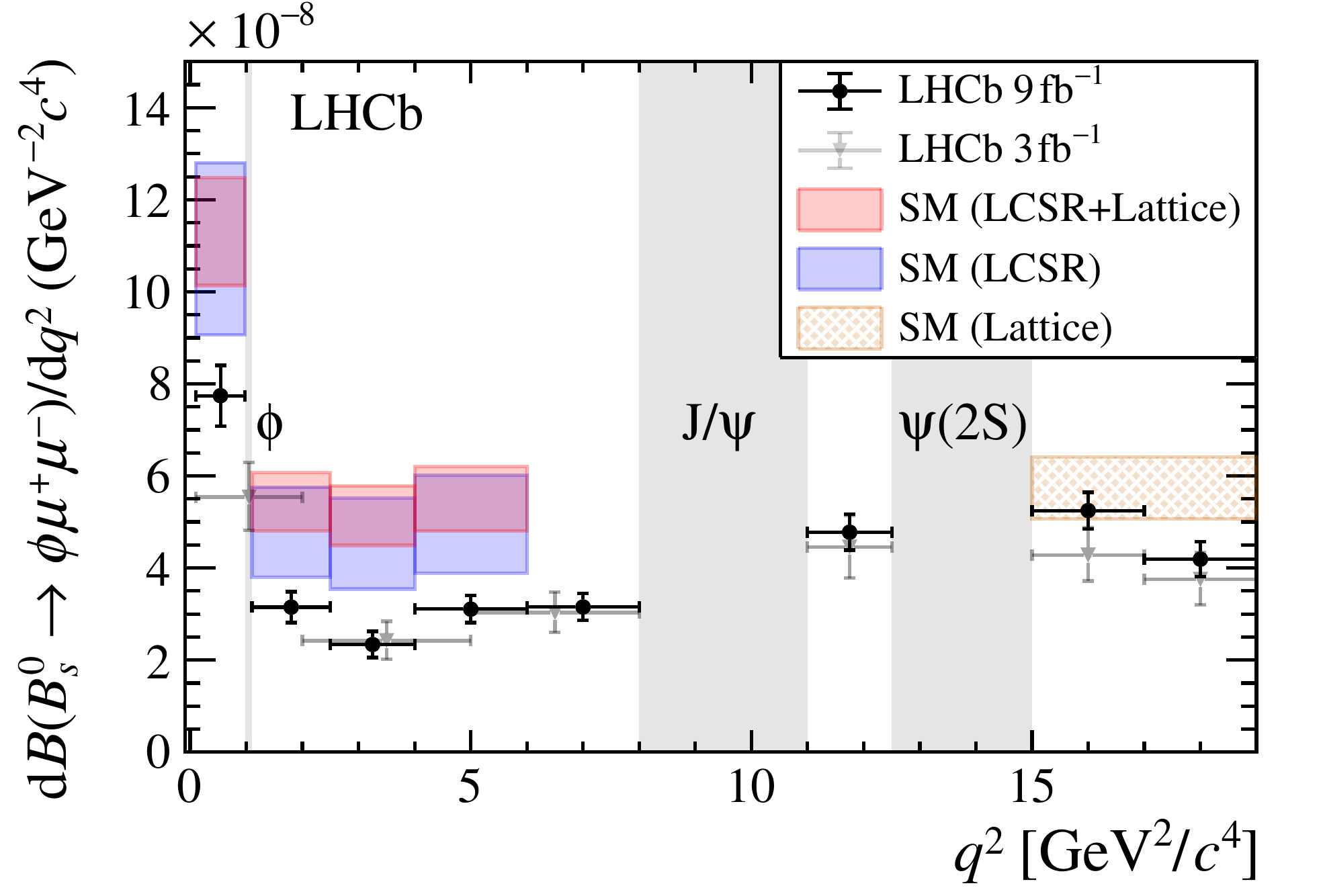}
    \caption{Differential branching fraction ${\rm d}{\mathcal B}(\decay{\Bs}{\phi\mumu})/{\rm d}q^2$, overlaid with SM predictions using Light Cone Sum Rules~\cite{Altmannshofer:2014rta,Straub:2015ica,Straub:2018kue} at low \qsq\ and Lattice calculations~\cite{Horgan:2013pva,Horgan:2015vla} at high \qsq.
    The results from the LHCb 3\invfb analysis~\cite{LHCb-PAPER-2015-023,LHCb-PAPER-2020-046} are shown with gray markers. }
    \label{fig:results}
\end{figure}

The resulting relative and total branching fractions are given in Table~\ref{tab:results}. 
In addition, the differential branching fraction is shown in Fig.~\ref{fig:results}, 
overlaid with SM predictions. These predictions are based on form factor calculations using Light Cone Sum Rules (LCSRs)~\cite{Altmannshofer:2014rta,Straub:2015ica} at low \qsq\ and Lattice QCD~(LQCD)~\cite{Horgan:2013pva,Horgan:2015vla} at high \qsq, which are implemented in the {\sc{flavio}}~software package~\cite{Straub:2018kue}. 
In  the \qsq region between $1.1$ and $6.0\gevgevcccc$, the measured branching fraction of $(2.88 \pm 0.22)\times 10^{-8}\gev^{-2}c^4$, lies $3.6\,\sigma$ below a precise SM prediction of $(5.37\pm0.66)\times 10^{-8}\gev^{-2}c^4$ which uses both LCSR and LQCD calculations. A less precise SM prediction of $(4.77\pm1.01)\times 10^{-8}\gev^{-2}c^4$ based on LCSRs alone lies $1.8\,\sigma$ above the measurement.
To determine the total branching fraction, 
the branching fractions of the individual \qsq\ intervals are summed
and corrected for the vetoed \qsq regions using $\epsilon_{\qsq\,\rm veto}=(65.47\pm 0.27)\,\%$. This efficiency is determined using SM simulation, and its uncertainty originates from the comparison of form factors from Ref.~\cite{Straub:2015ica} and Ref.~\cite{Ball:2004rg}.
The resulting 
branching fractions are
\begin{align*}
\frac{{\mathcal B}(\decay{\Bs}{\phi\mumu})}{{\mathcal B}(\decay{\Bs}{\jpsi\phi})}    &= (8.00 \pm 0.21 \pm 0.16 \pm 0.03)\times 10^{-4}~,\\
{\mathcal B}(\decay{\Bs}{\phi\mumu})    &=   (8.14 \pm 0.21 \pm 0.16 \pm 0.03 \pm 0.39)\times 10^{-7}, \nonumber
\end{align*}
where the uncertainties are, in order, statistical, systematic, from the extrapolation to the full \qsq\ region, and for the absolute branching fraction, from the branching fraction of the normalization mode. 
\begin{table}
\centering
\caption{
Differential ${\rm d}{\mathcal B}(\decay{\Bs}{\phi\mumu})/{\rm d}\qsq$ branching fraction, both relative to the normalization mode and absolute, in intervals of \qsq. The uncertainties are, in order, statistical, systematic, and due to the uncertainty on the branching fraction of the normalization mode.}
\label{tab:results}
\renewcommand*{\arraystretch}{1.25}
\begin{tabular}{rcc}\hline\noalign{\smallskip}
    $q^2$ interval & 
    ${\rm d} {\mathcal B}(\decay{\Bs}{\phi\mumu})/{\mathcal B}(\decay{\Bs}{\jpsi\phi}){\rm d}q^2$ & ${{\rm d} {\mathcal B}(\decay{\Bs}{\phi\mumu})/{\rm d}q^2}$ \\
    $[\gevgevcccc]$ & $[10^{-5}\gev^{-2}c^{4}]$ & $[10^{-8}\gev^{-2}c^{4}]$ \\
    \noalign{\smallskip}\hline\hline
$0.1\text{--}0.98$  &  $7.61 \pm 0.52 \pm 0.12$  & $7.74 \pm 0.53 \pm 0.12 \pm 0.37$ \\
$1.1\text{--}2.5$  	&  $3.09 \pm 0.29 \pm 0.07$  & $3.15 \pm 0.29 \pm 0.07 \pm 0.15$ \\
$2.5\text{--}4.0$  	&  $2.30 \pm 0.25 \pm 0.05$  & $2.34 \pm 0.26 \pm 0.05 \pm 0.11$ \\
$4.0\text{--}6.0$  	&  $3.05 \pm 0.24 \pm 0.06$  & $3.11 \pm 0.24 \pm 0.06 \pm 0.15$ \\
$6.0\text{--}8.0$  	&  $3.10 \pm 0.23 \pm 0.06$  & $3.15 \pm 0.24 \pm 0.06 \pm 0.15$ \\
$11.0\text{--}12.5$ &  $4.69 \pm 0.30 \pm 0.07$  & $4.78 \pm 0.30 \pm 0.08 \pm 0.23$ \\
$15.0\text{--}17.0$ &  $5.15 \pm 0.28 \pm 0.10$  & $5.25 \pm 0.29 \pm 0.10 \pm 0.25$ \\
$17.0\text{--}19.0$ &  $4.12 \pm 0.29 \pm 0.12$  & $4.19 \pm 0.29 \pm 0.12 \pm 0.20$ \\
\hline
$1.1\text{--}6.0$  	&  $2.83 \pm 0.15 \pm 0.05$  & $2.88 \pm 0.15 \pm 0.05 \pm 0.14$ \\
$15.0\text{--}19.0$ &  $4.55 \pm 0.20 \pm 0.11$  & $4.63 \pm 0.20 \pm 0.11 \pm 0.22$ \\
\hline\end{tabular}
\end{table}

The $\decay{\Bs}{\ftwoprime\mumu}$ decay is searched for using the combined \qsq\ region $[0.1,0.98] \cup [1.1,8.0] \cup [11.0,12.5]\gevgevcccc$. 
The branching fraction of the signal decay is determined relative to the $\decay{\Bs}{\jpsi\phi}$ normalization mode, 
according to
\begin{align}
    \frac{{\mathcal B}(\decay{\Bs}{\ftwoprime\mumu})}{
    {\mathcal B}(\decay{\Bs}{\jpsi\phi})} &= {\mathcal B}(\decay{\jpsi}{\mumu})\times\frac{{\mathcal B}(\decay{\phi}{\Kp\Km})}{{\mathcal B}(\decay{\ftwoprime}{\Kp\Km})}\times \frac{N_{\ftwoprime\mumu}}{N_{\jpsi\phi}}\times \frac{\epsilon_{\jpsi\phi}}{\epsilon_{\ftwoprime\mumu}},\label{eq:diffbf_f2}
\end{align}
where the ratio of branching fractions ${\mathcal B}(\decay{\phi}{\Kp\Km})/{\mathcal B}(\decay{\ftwoprime}{\Kp\Km})=1.123\pm 0.030$~\cite{pdg2020} is used. 
To separate the \ftwoprime\ signal from S- and P-wave contributions to the wide $m(\Kp\Km)$ mass window, 
a two-dimensional fit to the $m(\Kp\Km\mumu)$ and $m(\Kp\Km)$ distributions is performed. 
The $\decay{\Bs}{\ftwoprime\mumu}$ signal decay is modeled in $m(\Kp\Km\mumu)$ 
using the sum of two Gaussian functions with a power-law tail towards upper and lower mass,
and in $m(\Kp\Km)$ using a relativistic spin-2 Breit--Wigner function.
 The model parameters are determined from data using fits to the $\decay{\Bs}{\jpsi\ftwoprime}$ control mode and are fixed for the signal mode. 
Contributions from the S-wave and P-wave resonances, \eg\ the $\phi$ and the $\phi(1680)$ mesons,
are combined and described with a linear function in $m(\Kp\Km)$ and use the same model as the signal in $m(\Kp\Km\mumu)$. 
Interference effects are neglected as these were found to be small in the study of \mbox{$\decay{\Bs}{\jpsi\Kp\Km}$} decays in Ref.~\cite{LHCb-PAPER-2017-008}. 
The combinatorial background is modeled using an exponential function in both the reconstructed \Bs\ mass and the mass of the dikaon system. 
Background from $\decay{\Bd}{\Kp\pim\mumu}$ and $\decay{\Lb}{p\Km\mumu}$ decays is found to be non-negligible in the wide $m(\Kp\Km)$ window.
These background components are included in the fit model, with their yields constrained to the expected values and line shapes determined on simulated events. 

\begin{figure}
    \centering
    \includegraphics[width=7.5cm]{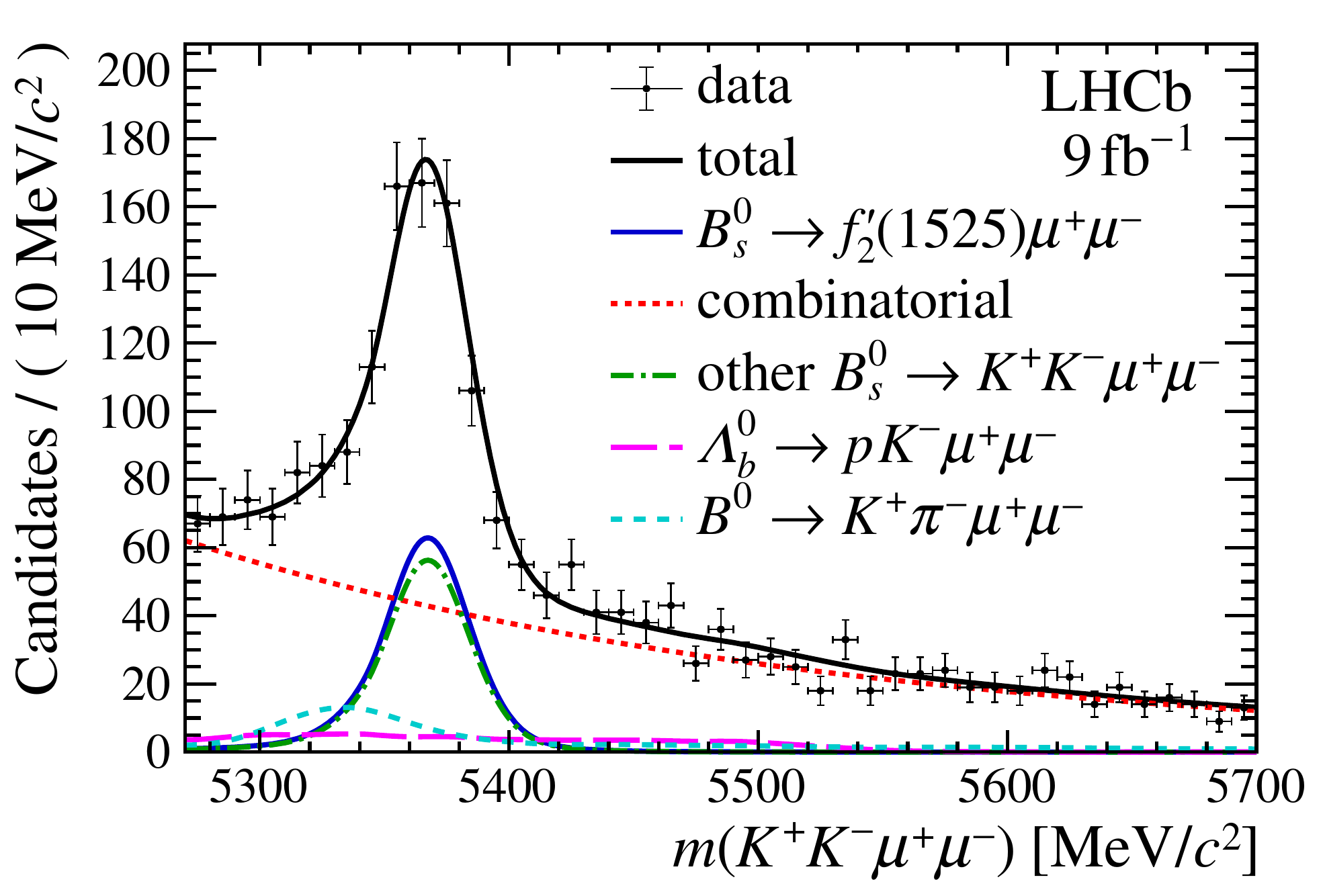}\hspace*{0.5cm}
    \includegraphics[width=7.5cm]{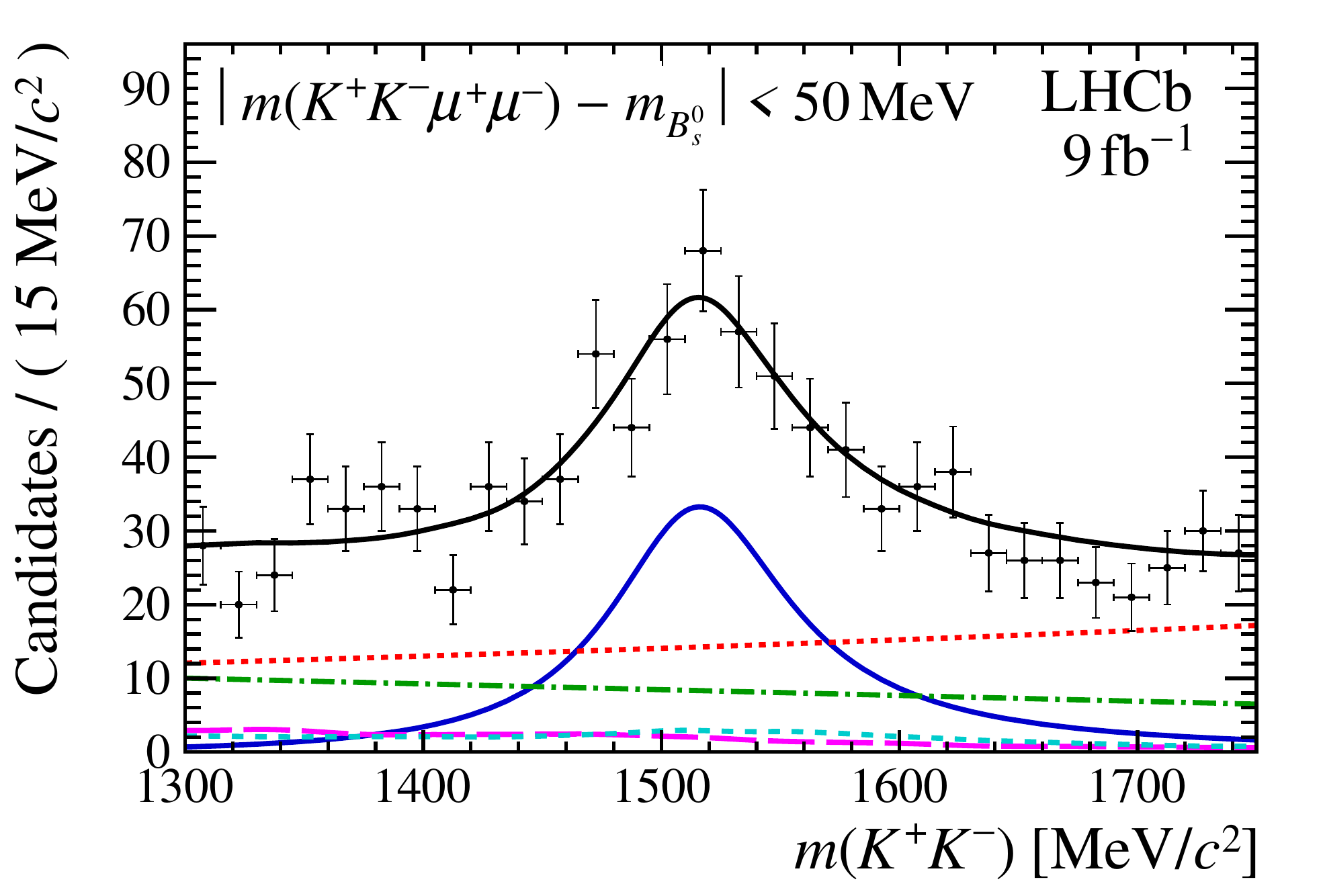}
    \caption{
    Reconstructed invariant mass of (left) the $\Kp\Km\mup\mun$ system and (right) the $\Kp\Km$ system for $\decay{\Bs}{\ftwoprime\mumu}$ candidates, overlaid with the fit projections. 
    The $m(\Kp\Km)$ distribution is shown in the \Bs\ signal region $\pm 50\mevcc$ around the known \Bs\ mass. 
    }
    \label{fig:ftwoprojections}
\end{figure}

The branching fraction of the $\decay{\Bs}{\ftwoprime\mumu}$ decay is determined using a simultaneous fit to the three data samples. 
The branching fraction of the signal and the S- and  P-wave contributions are shared between the data samples. 
From this fit, the signal yields, $N_{\ftwoprime\mumu}$, are found to be \mbox{$62\pm 8$}, \mbox{$67\pm 8$}, and \mbox{$161\pm 20$}  for the different data-taking periods.
Figure~\ref{fig:ftwoprojections} shows the $m(\Kp\Km\mumu)$ and $m(\Kp\Km)$ mass distributions, where the latter is shown  within $50\mevcc$ of the known \Bs\ mass~\cite{pdg2020}, overlaid with the fit projections.
 The significance of the signal is determined using Wilks' theorem~\cite{Wilks:1938dza}, comparing the log--likelihood with and without the signal component. 
The $\decay{\Bs}{\ftwoprime\mumu}$ decay is observed with a statistical significance of $9\,\sigma$. Systematic effects on the significance due to the choice of fit model are negligible. 

The dominant systematic uncertainties on the relative branching fraction of the $\decay{\Bs}{\ftwoprime\mumu}$ decay originate from the uncertainty of the branching fraction ratio ${\mathcal B}(\decay{\phi}{\Kp\Km})/{\mathcal B}(\decay{\ftwoprime}{\Kp\Km})$ ($0.04\times10^{-7}$), the modeling of the parameters of the Breit--Wigner function describing the $\ftwoprime$ resonance, and the simplified fit model for the $m(\Kp\Km)$ distribution ($0.03\times10^{-7}$). 
The effect of the simplified fit model 
is evaluated using pseudoexperiments, in which events are generated using the amplitude model in Ref.~\cite{LHCb-PAPER-2017-008} and fit with the default model. 
The observed difference in the determined yield is taken as a systematic uncertainty. 
Further details on the systematic uncertainties associated with $\mathcal{B}(\decay{\Bs}{\ftwoprime\mumu})$ are given in the Supplemental Material. 

The fraction of signal events within the considered \qsq\ region is calculated using the \qsq-differential distribution in Ref.~\cite{Rajeev:2020aut} and found to be $\epsilon_{\qsq\,\rm veto}=(73.8\pm 2.8)\,\%$. 
Accounting for this factor, 
the relative and total branching fractions are determined to be 
\begin{align*}
\frac{{\mathcal B}(\decay{\Bs}{\ftwoprime\mumu})}{{\mathcal B}(\decay{\Bs}{\jpsi\phi})}    &= (1.55 \pm 0.19 \pm 0.06 \pm 0.06)\times 10^{-4}~,\\
{\mathcal B}(\decay{\Bs}{\ftwoprime\mumu})    &=   (1.57 \pm 0.19 \pm 0.06 \pm 0.06 \pm 0.08)\times 10^{-7}~, \nonumber
\end{align*}
where the given uncertainties are, in order, statistical, systematic, from the extrapolation to the full \qsq range and, for the absolute branching fraction, from the uncertainty on the branching fraction of the normalization mode. 
The total $\decay{\Bs}{\ftwoprime\mumu}$ branching fraction is found to be 
in agreement with
 SM predictions~\cite{Li:2010ra,Zuo:2021kui,Rajeev:2020aut}. 

In summary, the most precise measurement of the branching fraction of the rare $\decay{\Bs}{\phi\mumu}$ decay is presented, using LHCb data corresponding to an integrated luminosity of $9\invfb$.
Consistent with earlier measurements~\cite{LHCb-PAPER-2015-023,LHCb-PAPER-2013-017}, the data are found to lie below SM expectations. In the \qsq\ region between $1.1$ and $6.0\gevgevcccc$ the measurement deviates by $3.6\,\sigma$ with respect to a precise SM prediction~\cite{Altmannshofer:2014rta,Straub:2015ica,Straub:2018kue,Horgan:2013pva,Horgan:2015vla}.
These results supersede, and are consistent with, those of  Refs.~\cite{LHCb-PAPER-2015-023,LHCb-PAPER-2013-017}.
In addition, the first observation of the rare $\decay{\Bs}{\ftwoprime\mumu}$ decay is reported with a statistical significance of nine standard deviations and the resulting branching fraction is found to be in agreement with 
SM predictions~\cite{Li:2010ra,Zuo:2021kui,Rajeev:2020aut}.

\section*{Acknowledgements}
%
%
\noindent We express our gratitude to our colleagues in the CERN
accelerator departments for the excellent performance of the LHC. We
thank the technical and administrative staff at the LHCb
institutes.
We acknowledge support from CERN and from the national agencies:
CAPES, CNPq, FAPERJ and FINEP (Brazil); 
MOST and NSFC (China); 
CNRS/IN2P3 (France); 
BMBF, DFG and MPG (Germany); 
INFN (Italy); 
NWO (Netherlands); 
MNiSW and NCN (Poland); 
MEN/IFA (Romania); 
MSHE (Russia); 
MICINN (Spain); 
SNSF and SER (Switzerland); 
NASU (Ukraine); 
STFC (United Kingdom); 
DOE NP and NSF (USA).
We acknowledge the computing resources that are provided by CERN, IN2P3
(France), KIT and DESY (Germany), INFN (Italy), SURF (Netherlands),
PIC (Spain), GridPP (United Kingdom), RRCKI and Yandex
LLC (Russia), CSCS (Switzerland), IFIN-HH (Romania), CBPF (Brazil),
PL-GRID (Poland) and NERSC (USA).
We are indebted to the communities behind the multiple open-source
software packages on which we depend.
Individual groups or members have received support from
ARC and ARDC (Australia);
AvH Foundation (Germany);
EPLANET, Marie Sk\l{}odowska-Curie Actions and ERC (European Union);
A*MIDEX, ANR, IPhU and Labex P2IO, and R\'{e}gion Auvergne-Rh\^{o}ne-Alpes (France);
Key Research Program of Frontier Sciences of CAS, CAS PIFI, CAS CCEPP, 
Fundamental Research Funds for the Central Universities, 
and Sci. \& Tech. Program of Guangzhou (China);
RFBR, RSF and Yandex LLC (Russia);
GVA, XuntaGal and GENCAT (Spain);
the Leverhulme Trust, the Royal Society
 and UKRI (United Kingdom).

\clearpage

\pagenumbering{roman}

\section*{Supplemental material}

\subsection*{Supplemental figures}

\begin{figure}[h]
    \centering
    \includegraphics[width=10.5cm]{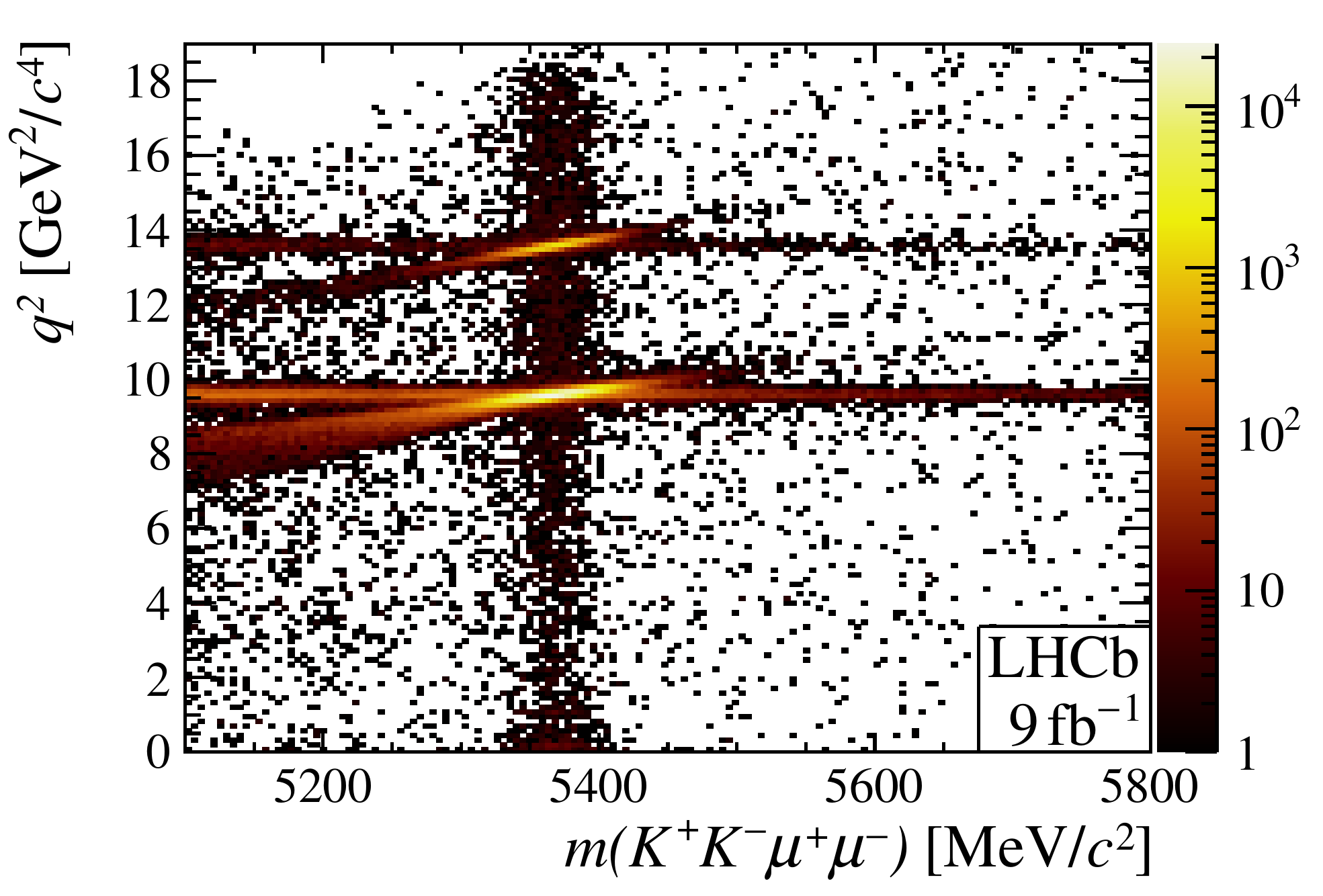}\\ 
    \includegraphics[width=10.5cm]{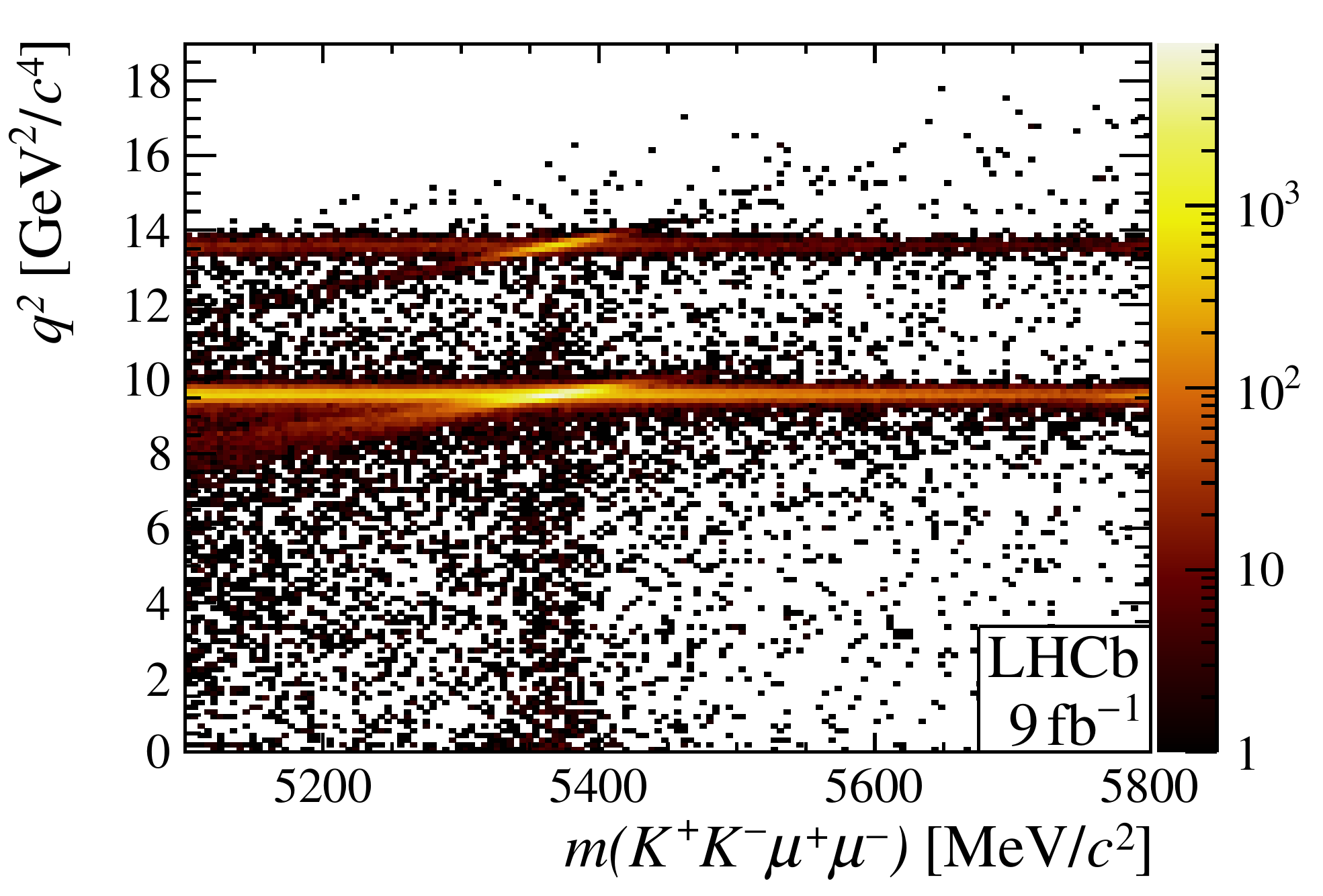}
    \caption{Invariant mass of the $\Kp\Km\mup\mun$ system versus \qsq\ for selected (top) $\decay{\Bs}{\phi\mumu}$ and  (bottom) $\decay{\Bs}{\ftwoprime\mumu}$  candidates across all data-taking periods.
    }
    \label{fig:mbsq2}
\end{figure}

Figure~\ref{fig:mbsq2} shows the $\Kp\Km\mumu$ invariant mass versus \qsq for selected (top) \mbox{\decay{\Bs}{\phi\mumu}} and (bottom) \mbox{\decay{\Bs}{\ftwoprime\mumu}}  candidates. 
The signal modes are clearly visible as a vertical band around the known mass of the \Bs meson.

\clearpage 

\begin{figure}[h]
    \centering
    \includegraphics[width=7.5cm]{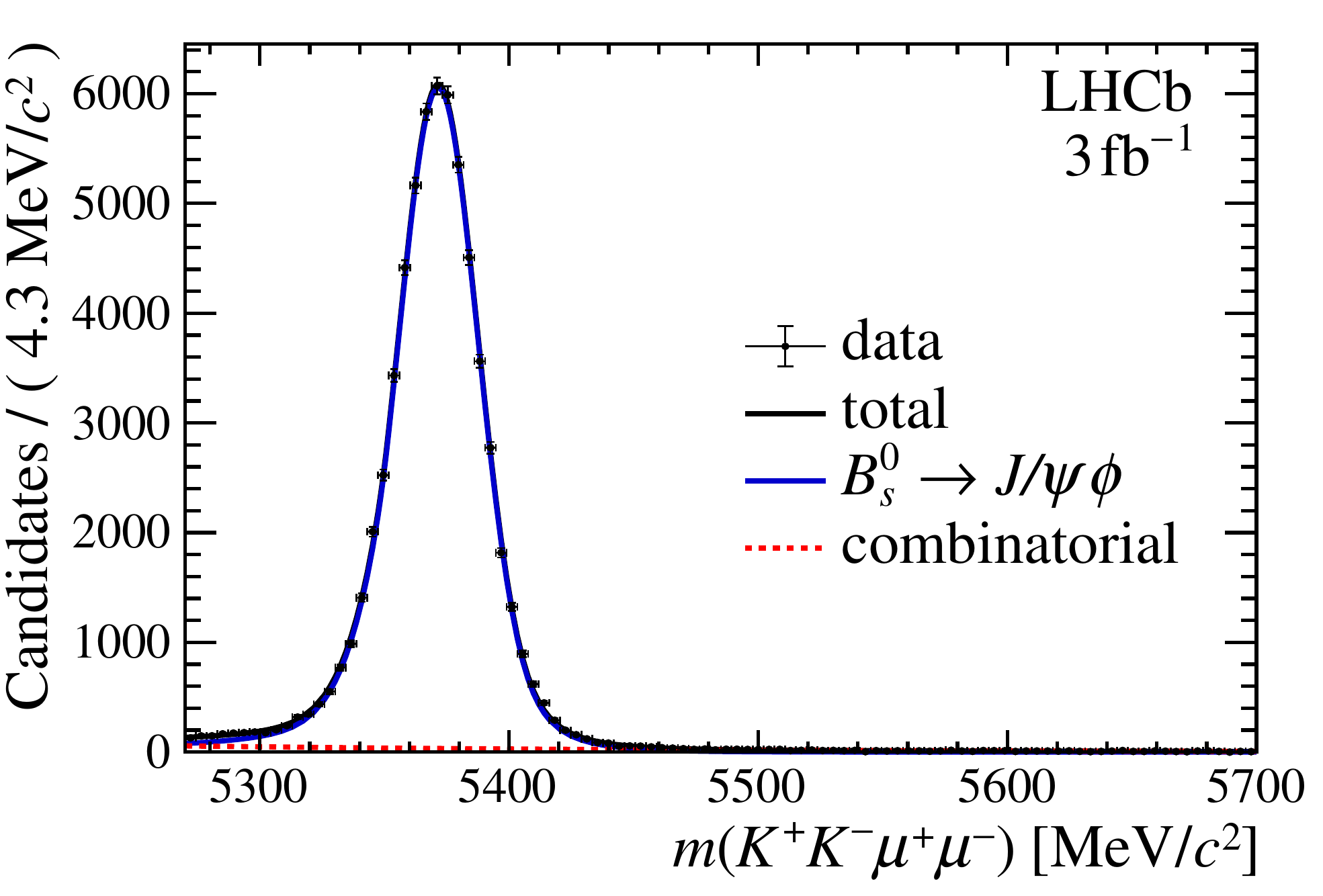}\hspace*{0.5cm}
    \includegraphics[width=7.5cm]{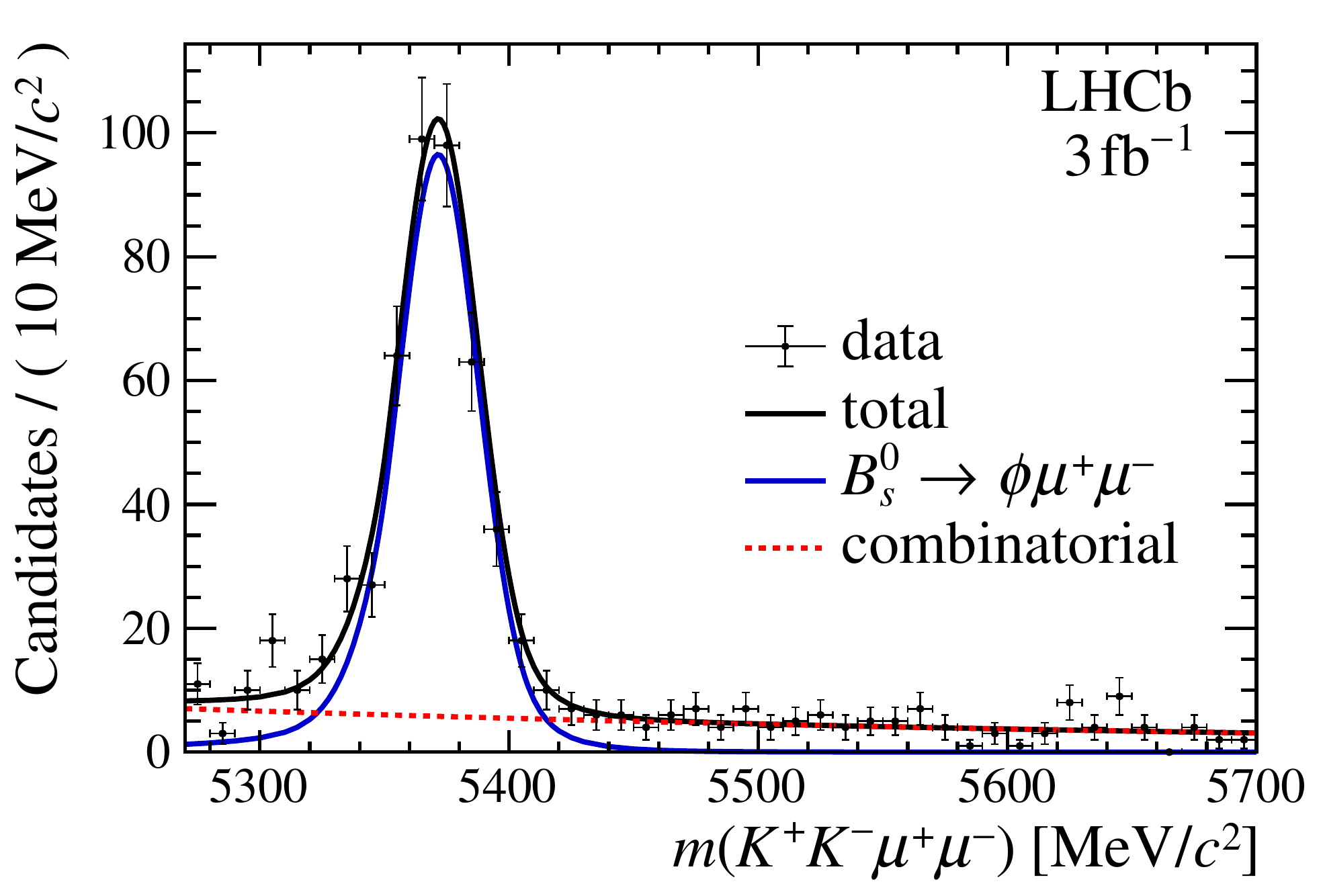}\\
    \includegraphics[width=7.5cm]{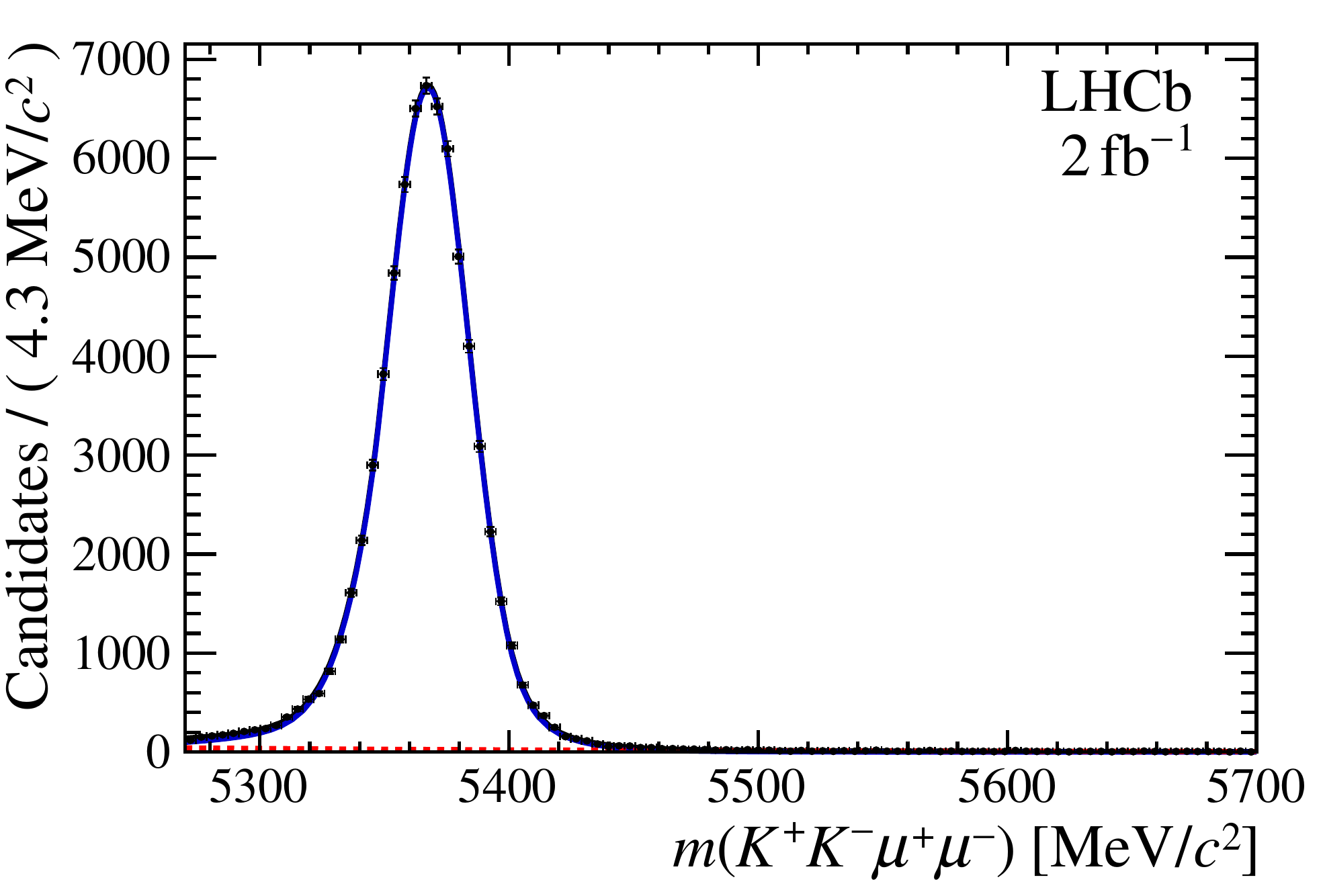}\hspace*{0.5cm}
    \includegraphics[width=7.5cm]{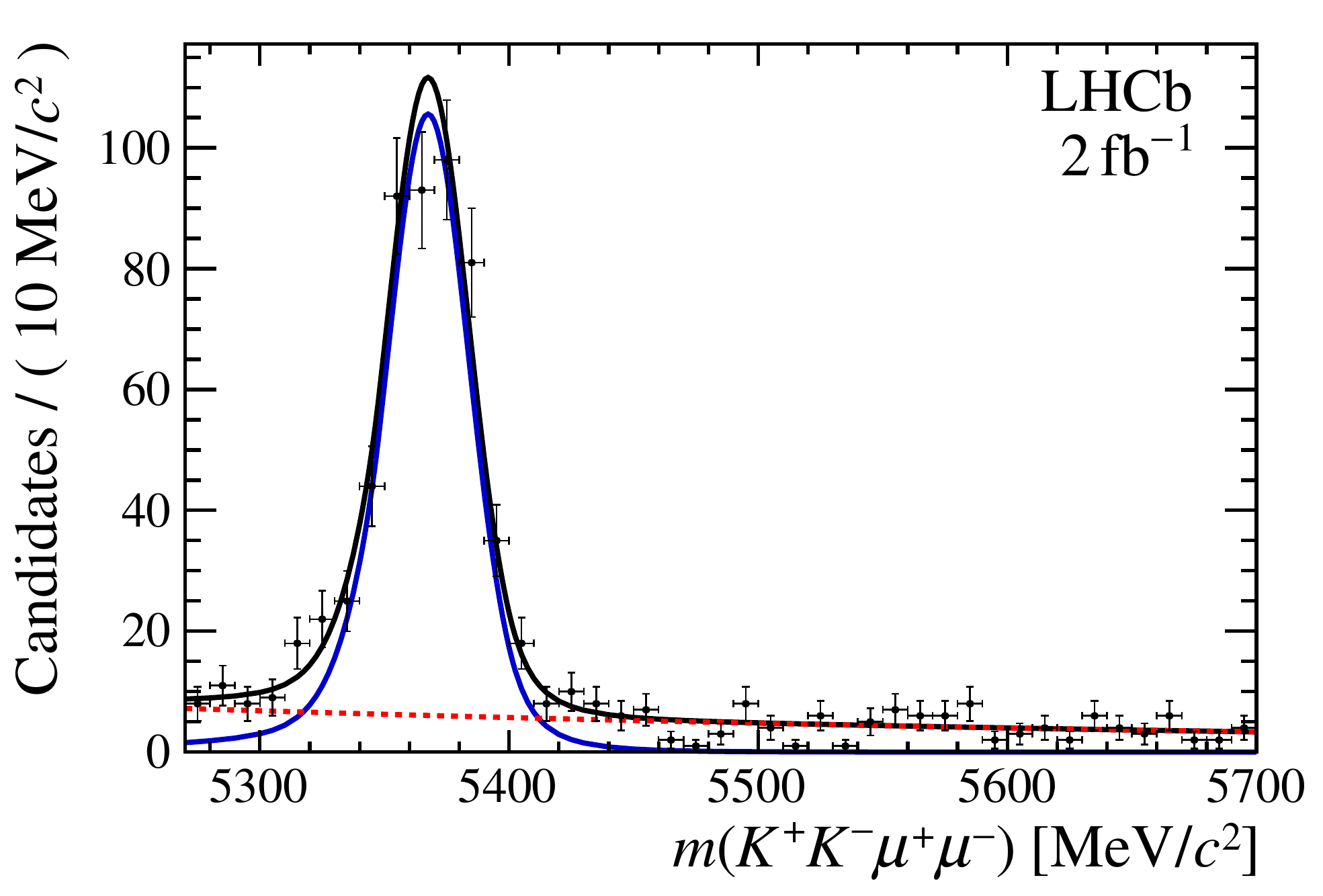}\\
    \includegraphics[width=7.5cm]{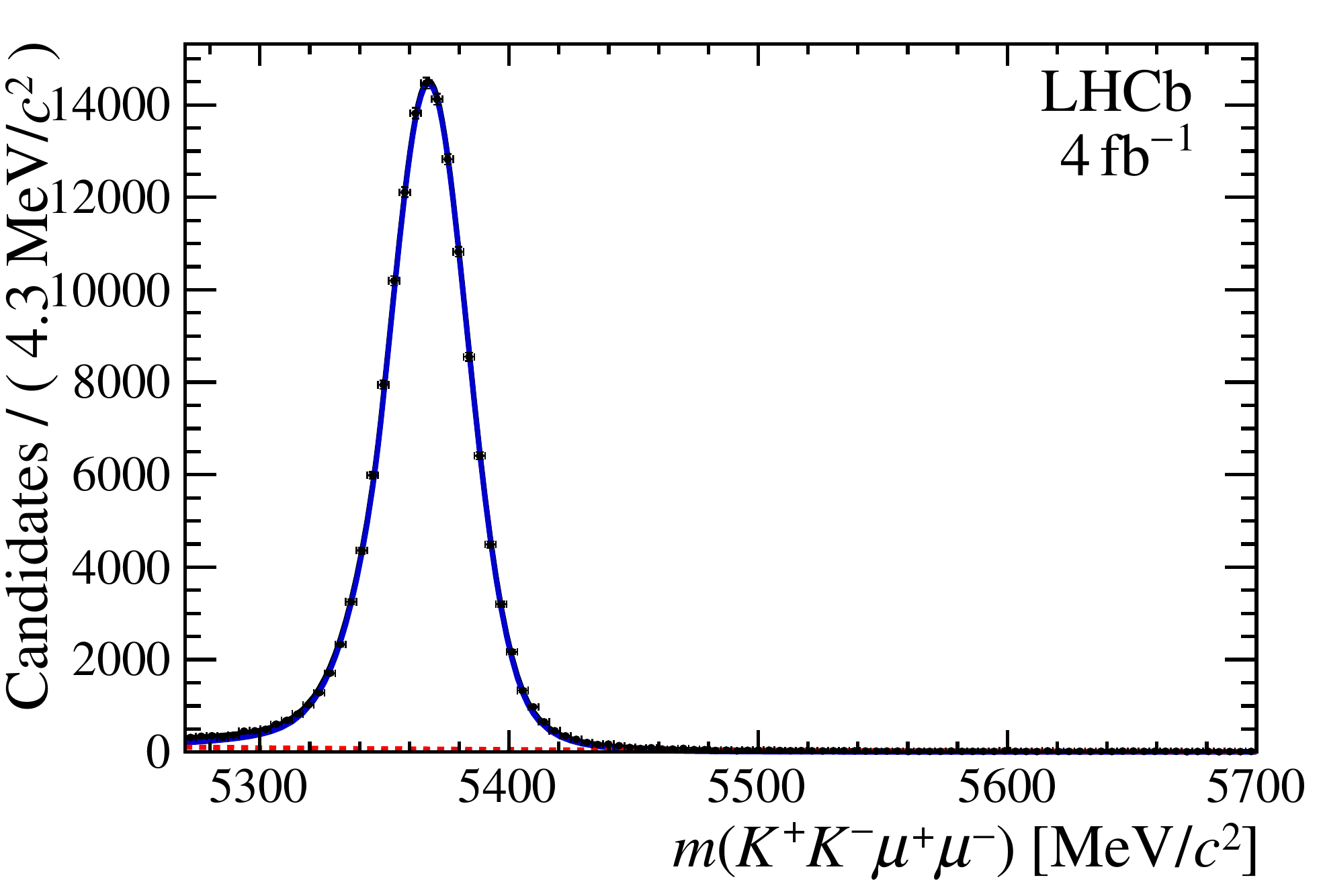}\hspace*{0.5cm}
    \includegraphics[width=7.5cm]{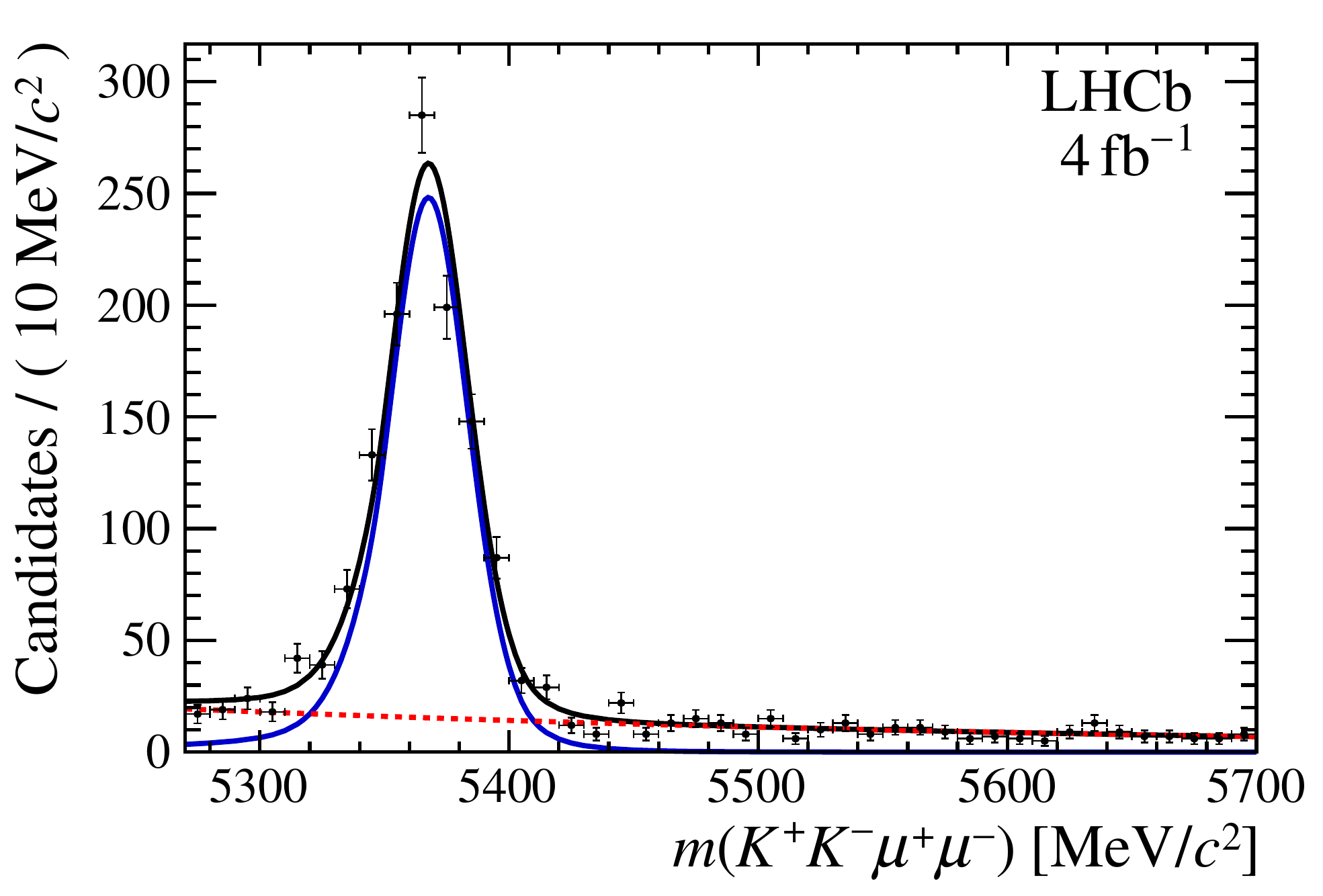}
    \caption{
Reconstructed invariant mass of the $\Kp\Km\mumu$ system 
       for
       (left) $\decay{\Bs}{\jpsi\phi}$ and  (right) $\decay{\Bs}{\phi\mumu}$ candidates, integrated over \qsq,  for the (top) 2011--2012, (middle) 2015--2016, and (bottom) 2017--2018 data-taking periods. The data are overlaid with 
       the fit projections.
     }
    \label{fig:phiprojections_separate}
\end{figure}

The  $\Kp\Km\mumu$ invariant mass  of the selected (left) \mbox{\decay{\Bs}{\jpsi\phi}} and (right) \mbox{\decay{\Bs}{\phi\mumu}} candidates, integrated over \qsq, is shown in Fig.~\ref{fig:phiprojections_separate} for the different data-taking periods.
The total 
fit projection
(black line) is overlaid on the data, along with the signal component (blue line) and background component describing combinatorial background (red dotted line).

\clearpage

\begin{figure}
    \centering
    \includegraphics[width=7.5cm]{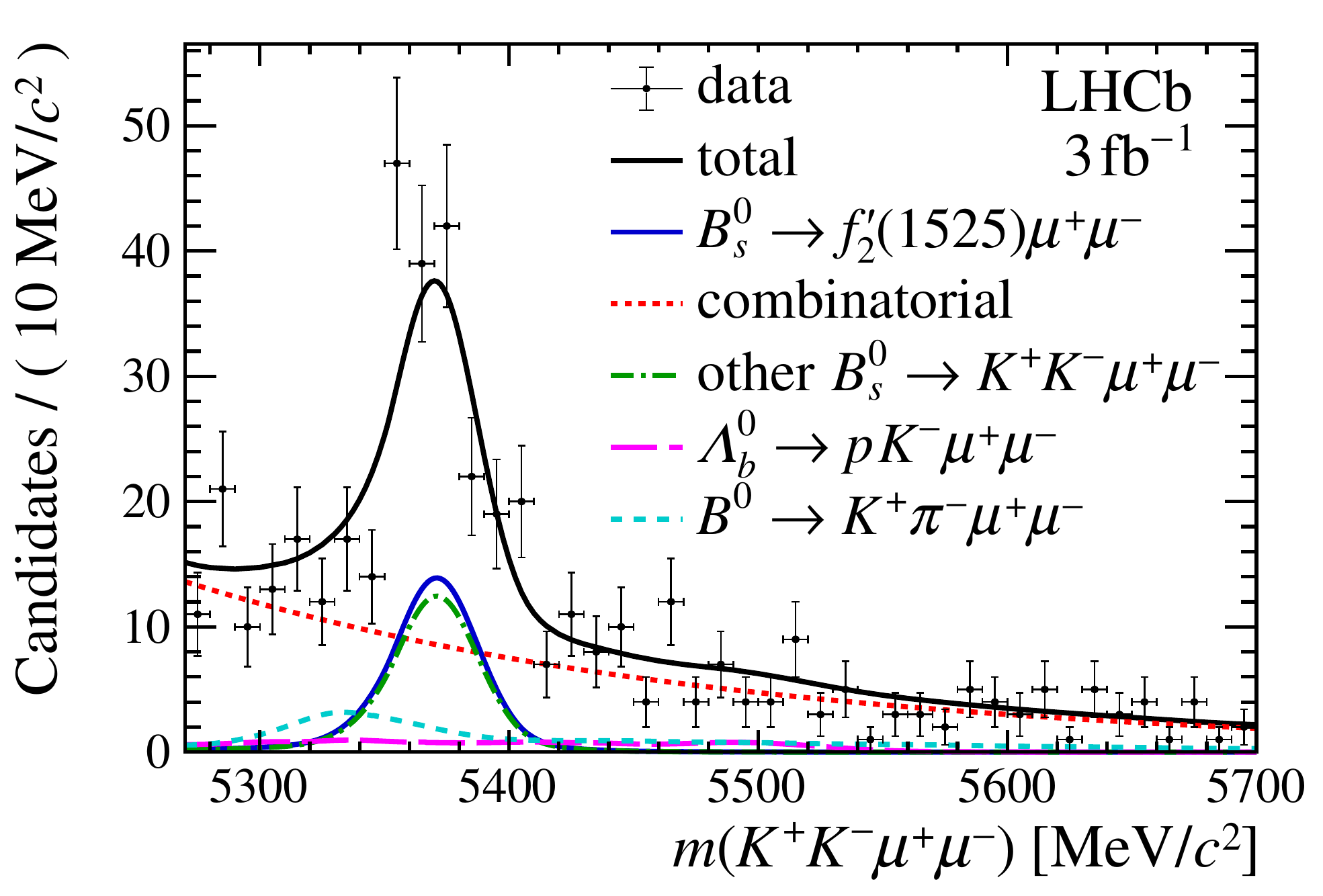}\hspace*{0.5cm}
    \includegraphics[width=7.5cm]{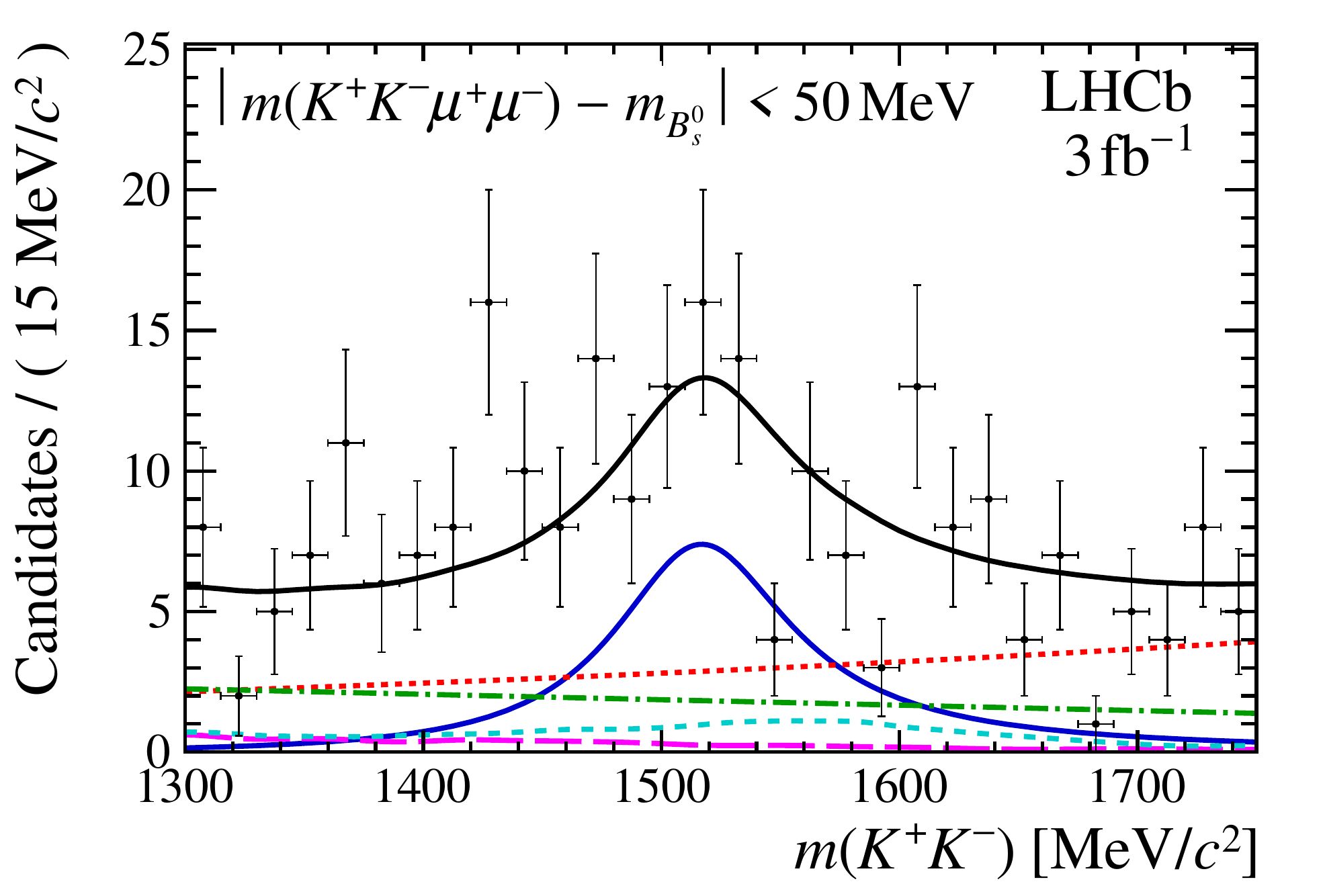}\\
    \includegraphics[width=7.5cm]{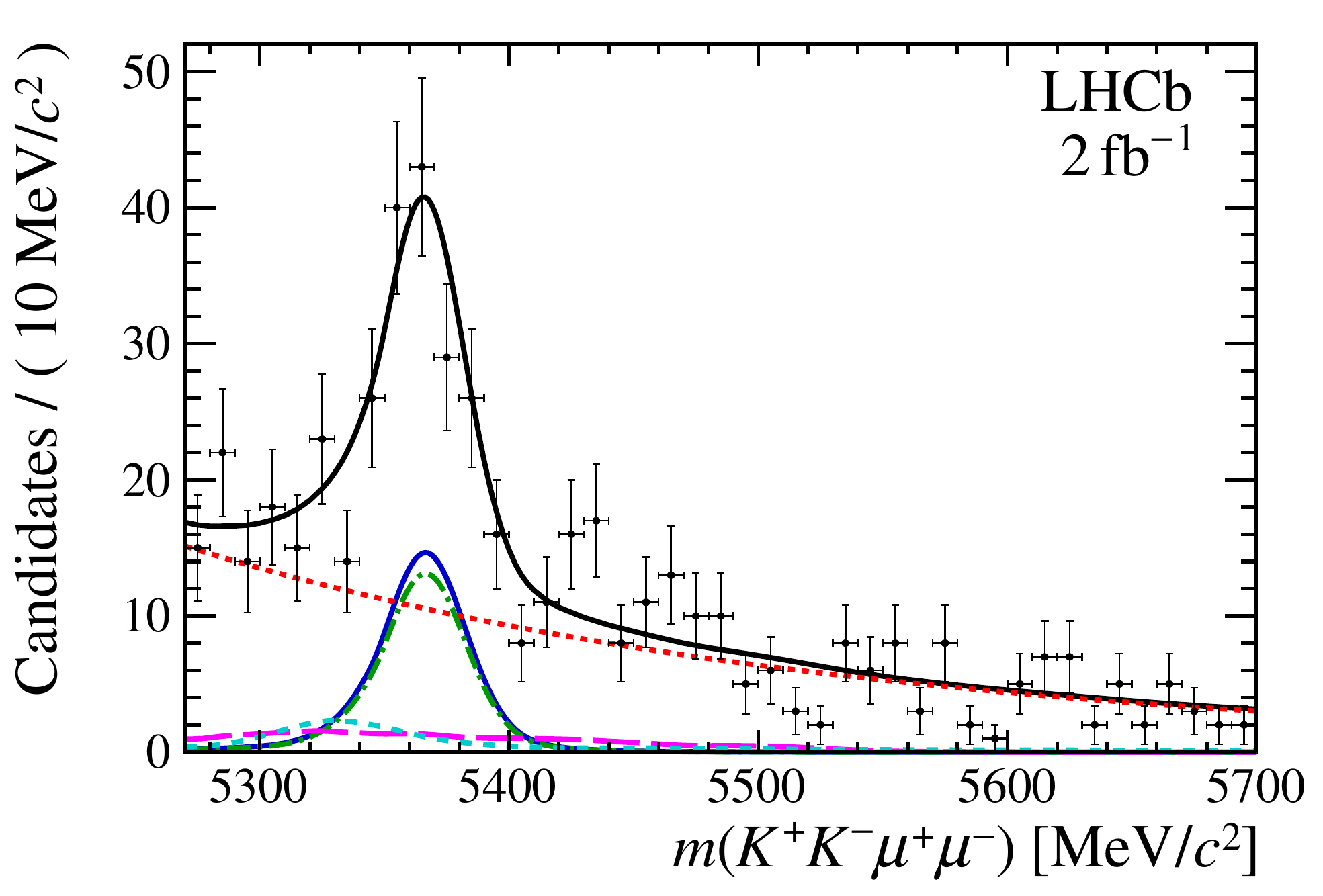}\hspace*{0.5cm}
    \includegraphics[width=7.5cm]{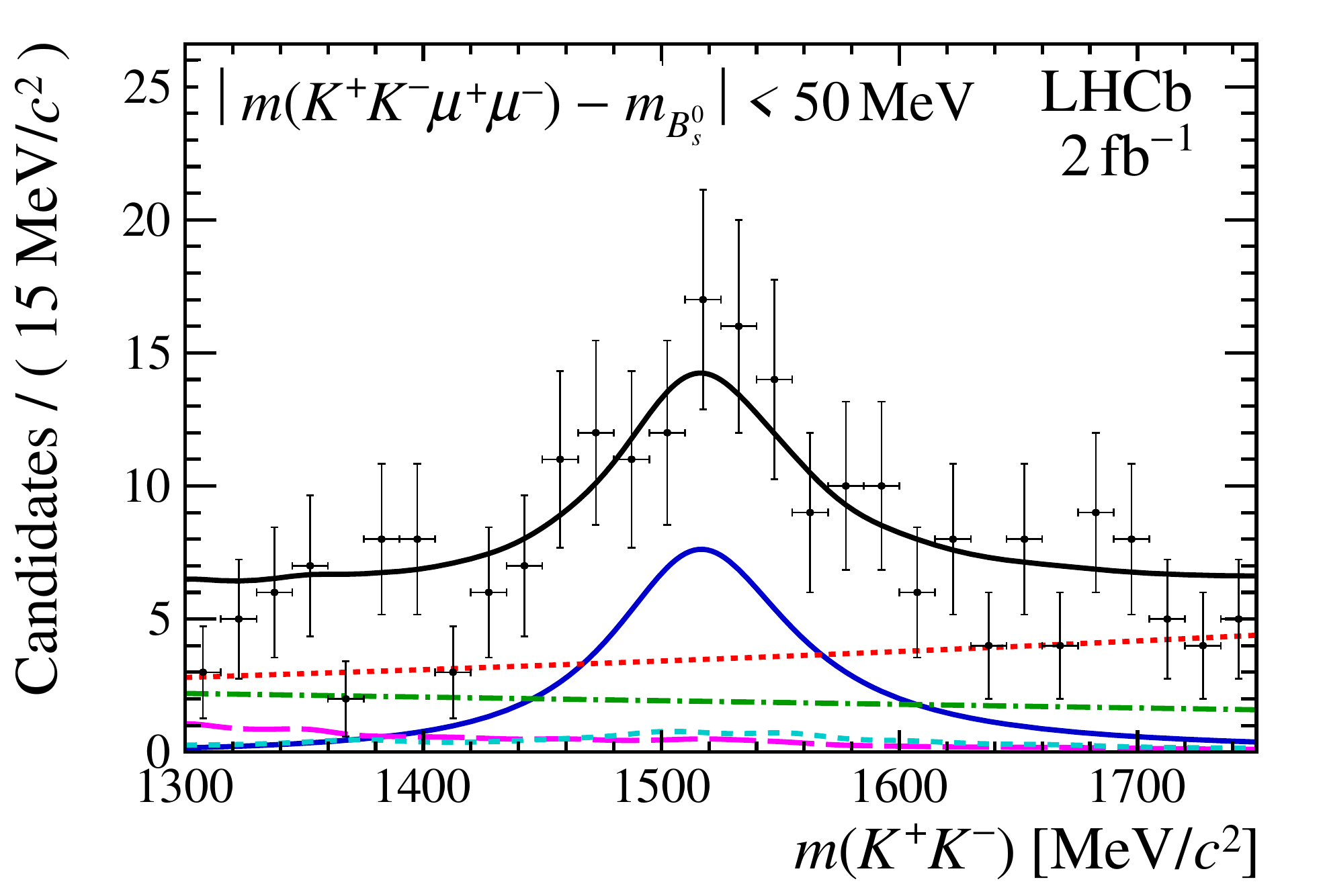}\\
    \includegraphics[width=7.5cm]{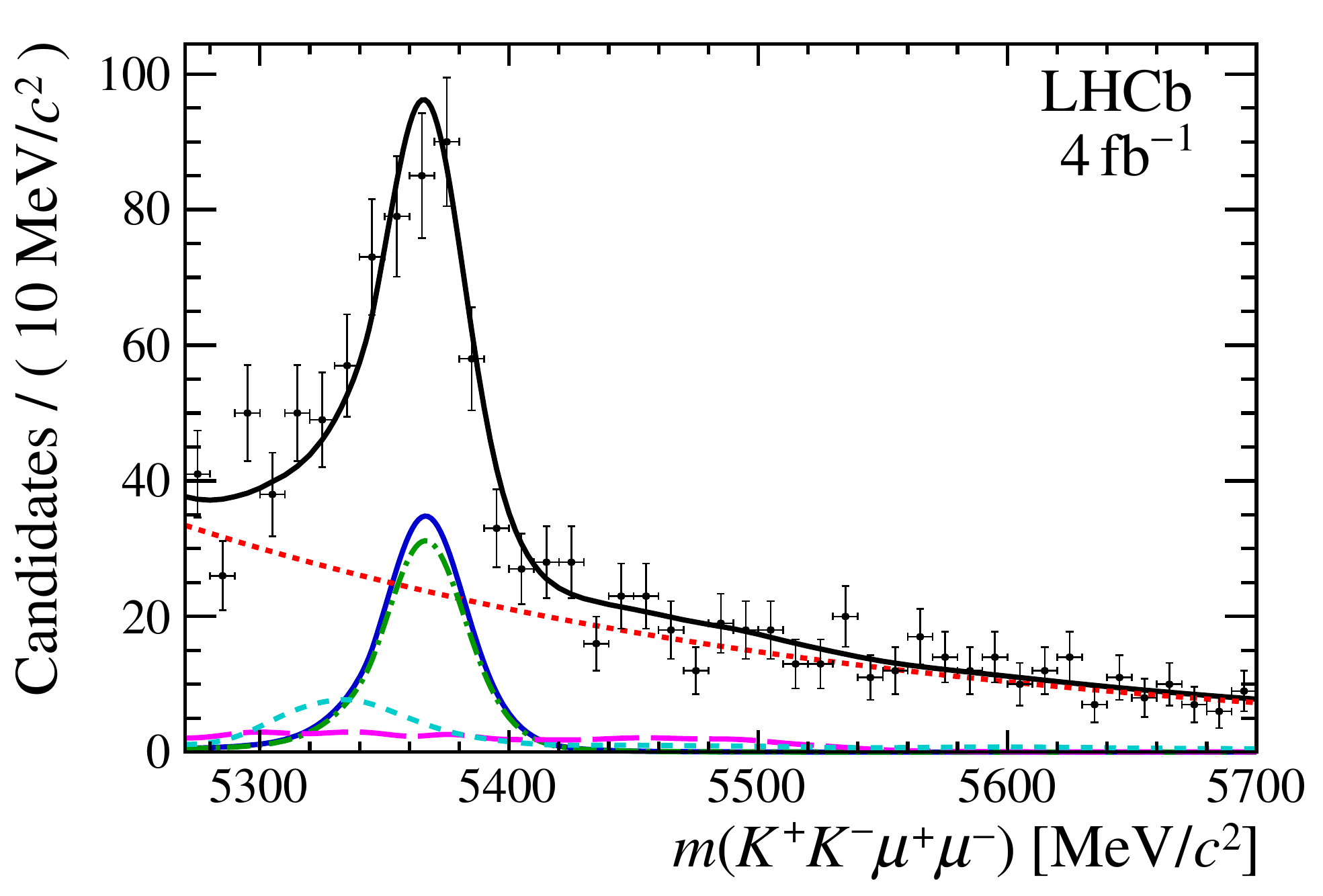}\hspace*{0.5cm}
    \includegraphics[width=7.5cm]{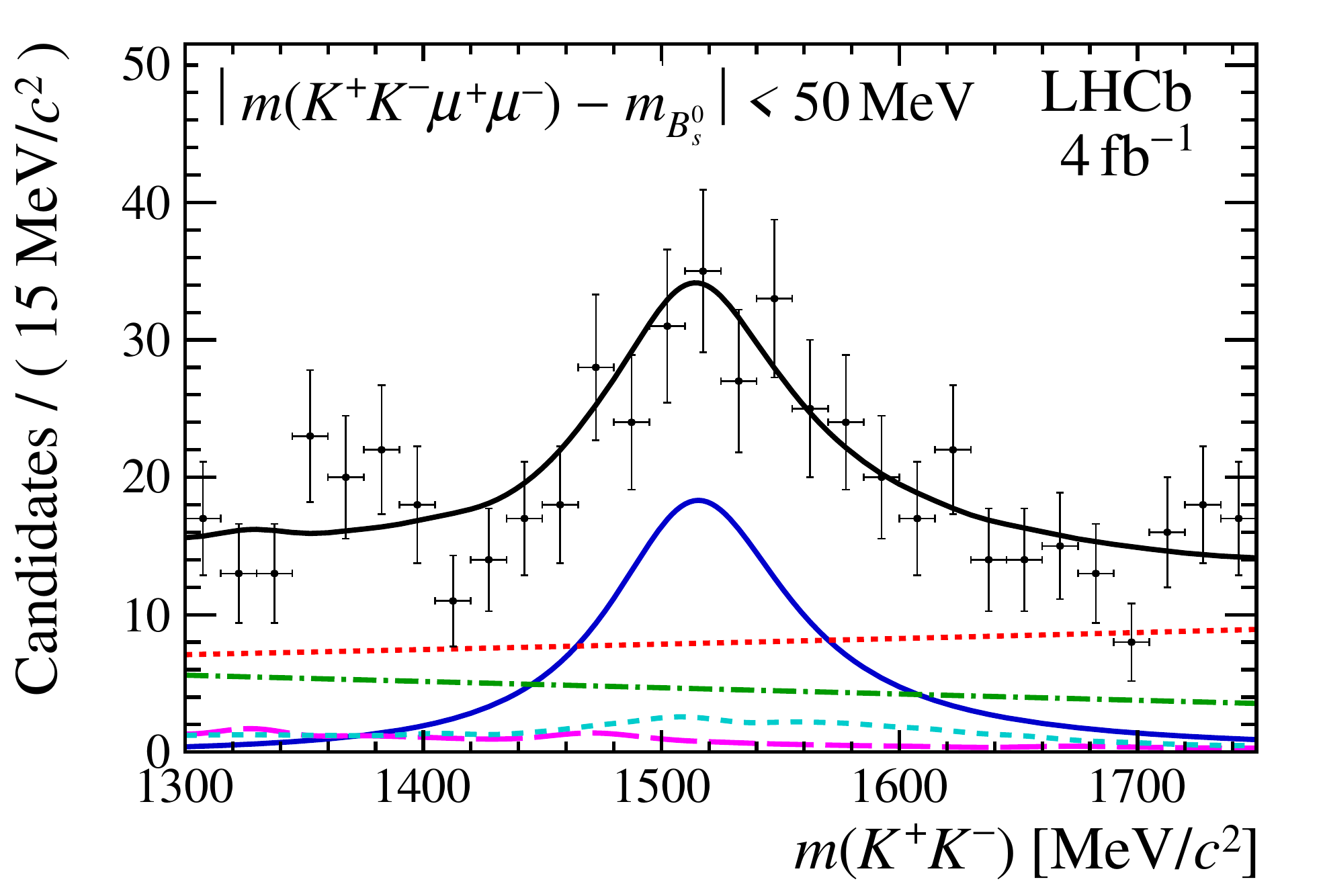}
    \caption{
    Reconstructed invariant mass of (left) the $\Kp\Km\mup\mun$ system and (right) the $\Kp\Km$ system for  $\decay{\Bs}{\ftwoprime\mumu}$ candidates for the (top) 2011--2012, (middle) 2015--2016, and (bottom) 2017--2018 data-taking periods. Distributions are overlaid with the
    fit projections.
   The $\Kp\Km$ distribution is shown in the \Bs signal region $\pm50\mev$ around the known \Bs mass. 
    }
    \label{fig:ftwoprojections_separate}
\end{figure}

Figure~\ref{fig:ftwoprojections_separate} shows the (left) $\Kp\Km\mumu$  and (right) $\Kp\Km$ invariant mass distributions of selected \mbox{\decay{\Bs}{\ftwoprime\mumu}} candidates for the different data-taking periods. 
The total fit projection
(black line) is overlaid on the data along with projections of individual fit components describing: the signal (blue line), other \mbox{\decay{\Bs}{\Kp\Km\mumu}} decays (green dash-dotted line), combinatorial background (red dotted line) and \Lb (magenta long dashed line) and \Bd (cyan medium size dashed line) decays.

\clearpage

\subsection*{Systematic uncertainties}

The systematic uncertainties associated with the measurement of the branching fractions of the $\decay{\Bs}{\phi\mumu}$ and $\decay{\Bs}{\ftwoprime\mumu}$ decays are summarized in Table~\ref{tab:systs_suppl}.  The \emph{Physics model} in Table~\ref{tab:systs_suppl} refers to the model used for the generation of $\decay{\Bs}{\phi\mumu}$ and $\decay{\Bs}{\ftwoprime\mumu}$ decays in simulation. 
Studied variations to the $\decay{\Bs}{\phi\mumu}$ physics model are detailed in the Letter. The model used for $\decay{\Bs}{\ftwoprime\mumu}$ decays accounts only for phase-space effects. 
The \qsq distribution is therefore not an exact description of data and is corrected to the predictions in Ref.~\cite{Rajeev:2020aut}. 
The difference in the relative efficiency with and without this correction is assigned as a systematic uncertainty.

The \emph{Residual background} in Table~\ref{tab:systs_suppl} refers to contamination of the signal modes from residual physical background. 
 Background contributions to $\decay{\Bs}{\phi\mumu}$ decays are neglected in the fit and a systematic uncertainty is assigned.
 For $\decay{\Bs}{\ftwoprime\mumu}$ decays, a systematic uncertainty is associated with the choice of lineshape used to describe background from $\decay{\Bd}{\Kp\pim\mumu}$ and $\decay{\Lb}{p\Km\mumu}$ decays in the fit. 

The systematic uncertainty associated with the $\decay{\Bs}{\phi\mumu}$ \emph{signal fit model} in Table~\ref{tab:systs_suppl} is obtained using an alternative description for the radiative tail of the \Bs\ meson. 
For the \decay{\Bs}{\ftwoprime\mumu} decay, the lineshape in $m(K^{+}K^{-})$  of both the non-signal \decay{\Bs}{K^{+}K^{-}\mumu} contributions and the signal decay are varied. For the \decay{\Bs}{K^{+}K^{-}\mumu} contributions, an alternative description
is taken from Ref.~\cite{LHCb-PAPER-2017-008}, as detailed in the Letter. 
For the \ftwoprime\ lineshape, the input values for the Blatt--Weisskopf barrier functions\cite{Blatt:1952ije} are varied, namely the barrier radius  of the \ftwoprime\ and \Bs\ mesons, along with the orbital angular momentum of the \Bs meson.

The \emph{simulation corrections} in Table~\ref{tab:systs_suppl} refer to the uncertainties associated with applied corrections to simulated events. 
Corrections are recalculated using alternative binning schemes or accounting for the finite statistics of the control modes used to derive the corrections. 
The uncertainty associated with small levels of mismodelling in distributions which are not directly corrected for in the default approach (e.g. tracking efficiencies) are indicated under \emph{residual mismodelling}.

\begin{table}[h]
\centering
\caption{Systematic uncertainties on the differential branching fraction \mbox{${\rm d}{\mathcal B}(\decay{\Bs}{\phi\mumu})/{\rm d}q^2$} and on the total branching fraction \mbox{${\mathcal B}(\decay{\Bs}{\ftwoprime\mumu})$}. Ranges indicate the variation across the \qsq\ intervals.
The uncertainty from the branching fraction of the normalization mode, ${\mathcal B}(\decay{\Bs}{\jpsi\phi)}$, is quoted separately.
}
\label{tab:systs_suppl}
\renewcommand*{\arraystretch}{1.25}
\begin{tabular}{lrr}\hline

Source & $\sigma_{\rm syst.}\bigl({\rm d}{\mathcal B}(\decay{\Bs}{\phi\mumu})/{\rm d}q^2\bigr)$ &  $\sigma_{\rm syst.}\bigl({\mathcal B} (\decay{\Bs}{\ftwoprime\mumu})\bigr)$ \\
 & $[10^{-8}\gev^{-2}c^4]$ & $[10^{-7}]$\\\hline\hline
Physics model             & $0.04\text{--}0.10$  & $0.02$ \\
Limited simulation sample & $0.02\text{--}0.07 $ & $0.01$ \\
Residual background   & $0.01\text{--}0.04$  & $0.01$ \\
Fit bias               & $0.00\text{--}0.03$  & $<0.01$\\
Signal fit model           & $0.00\text{--}0.01$  & $0.03$\\
Simulation corrections & $0.00\text{--}0.03$  & $0.01$\\
Residual mismodelling  & $0.00\text{--}0.02$ & $<0.01$\\
${\mathcal B}(\decay{\jpsi}{\mumu})$     & $0.01\text{--}0.04$ & $0.01$  \\
${\mathcal B}(\decay{\phi}{\Kp\Km})/{\mathcal B}(\decay{\ftwo}{\Kp\Km})$     & -- & $0.04$\\
\hline
Quadratic sum             & $0.05\text{--}0.12$ & $0.06$ \\
 \hline
 Normalization ${\mathcal B}(\decay{\Bs}{\jpsi\phi)}$ & $0.11\text{--}0.37$ & $0.07$ \\
\hline\end{tabular}
\end{table}

\clearpage

\addcontentsline{toc}{section}{References}

\ifx\mcitethebibliography\mciteundefinedmacro
\PackageError{LHCb.bst}{mciteplus.sty has not been loaded}
{This bibstyle requires the use of the mciteplus package.}\fi
\providecommand{\href}[2]{#2}

\newpage
\centerline
{\large\bf LHCb collaboration}
\begin
{flushleft}
\small
R.~Aaij$^{32}$,
C.~Abell{\'a}n~Beteta$^{50}$,
T.~Ackernley$^{60}$,
B.~Adeva$^{46}$,
M.~Adinolfi$^{54}$,
H.~Afsharnia$^{9}$,
C.A.~Aidala$^{86}$,
S.~Aiola$^{25}$,
Z.~Ajaltouni$^{9}$,
S.~Akar$^{65}$
% ...

{\footnotesize \it

% ...
$^{9}$Universit{\'e} Clermont Auvergne, CNRS/IN2P3, LPC, Clermont-Ferrand, France\\
% ...
$^{25}$INFN Sezione di Milano, Milano, Italy\\
% ...
$^{32}$Nikhef National Institute for Subatomic Physics, Amsterdam, Netherlands\\
% ...
$^{46}$Instituto Galego de F{\'\i}sica de Altas Enerx{\'\i}as (IGFAE), Universidade de Santiago de Compostela, Santiago de Compostela, Spain\\
% ...
$^{50}$Physik-Institut, Universit{\"a}t Z{\"u}rich, Z{\"u}rich, Switzerland\\
% ...
$^{54}$H.H. Wills Physics Laboratory, University of Bristol, Bristol, United Kingdom\\
% ...
$^{60}$Oliver Lodge Laboratory, University of Liverpool, Liverpool, United Kingdom\\
% ...
$^{65}$University of Cincinnati, Cincinnati, OH, United States\\
% ...
$^{68}$Syracuse University, Syracuse, NY, United States\\
% ...
$^{86}$University of Michigan, Ann Arbor, United States, associated to $^{68}$\\
\bigskip
% ...
\medskip
}
\end{flushleft}

\clearpage

\end{document}